\documentclass[preprint,aps,prd,nofootinbib,showpacs,showkeys,12pt]{revtex4}%
\usepackage{amssymb}
\usepackage{graphicx}
\usepackage{graphics}
\usepackage{epsfig}
\usepackage{amsmath}
\usepackage{amsfonts}

\begin{document}

\title{\textbf{{\Large Analysis of U$_A$(1) symmetry breaking and restoration effects on the scalar-pseudoscalar meson spectrum}}}

\author{\textbf{P. Costa}}
\email{pcosta@teor.fis.uc.pt}
\author{\textbf{M. C. Ruivo}}
\email{maria@teor.fis.uc.pt}
\author{\textbf{C. A. de Sousa}}
\email{celia@teor.fis.uc.pt}
\affiliation{Departamento de F\'{\i}sica, Universidade de Coimbra, P-3004-516 Coimbra, Portugal}
\author{\textbf{Yu. L. Kalinovsky}}
\email{kalinov@nusun.jinr.ru}
\affiliation{Laboratory of Information Technologies, Joint Institute for Nuclear Research,
Dubna, Russia}

%%%%%%%%%%%%%%%%%%%%%%%%%%%%%%%%%%%%%%%%%%%%%%%%%%%%%%%%%%%%%%%%%%%%%%%%%%%%%%%%%%%%%%%%%%%%%%%%%%%%%%%%%%%%%%%%%%%%%%%%%%%%%%%%%%%%%%%%%%%%%%%%%%%%%%%%%%%%%%%%%%
%%%%%%%%%%%%%%%%%%%%%%%%%%%%%%%%%%%%%%%%%%%%%%%%%%%%%%%%%%%%%%%%%%%%%%%%%%%%%%%%%%%%%%%%%%%%%%%%%%%%%%%%%%%%%%%%%%%%%%%%%%%%%%%%%%%%%%%%%%%%%%%%%%%%%%%%%%%%%%%%%%

\begin{abstract}
We explore patterns of effective restoration of the chiral U$_A$(1) symmetry using an extended  three-flavor Nambu-Jona-Lasinio model that incorporates explicitly the axial anomaly through the 't Hooft interaction, and assuming that the coefficient of the anomaly term is temperature and density dependent. The special case of explicit breaking of chiral symmetry without U$_A$(1) anomaly is also considered, since  this scenario can provide additional information allowing  to understand the interplay between the U$_A$(1) anomaly and (spontaneous) chiral symmetry breaking effects. The pseudoscalar and scalar sectors are  analyzed in detail bearing in mind the identification of chiral partners and the study of its convergence. We also concentrate on the behavior of the mixing angles that give us relevant information on the issue under discussion. 
In the region of temperatures (densities) studied, we do not observe signs indicating a full restoration of U(3)$\otimes$U(3)  symmetry as, for instance, the degeneracy of both $a_0$ and $f_0$ mesons with the pion.
As we work in a real world scenario ($m_u=m_d<<m_s$),  we only observe the return to symmetries of the classical QCD Lagrangian in the non-strange sector.
\end{abstract}
\pacs{11.10.Wx, 11.30.Rd, 14.40.Aq, 24.85.+p} 
\keywords{NJL model, U$_A$(1) symmetry, Restoration of chiral and axial symmetries, Chiral partners }
\maketitle

%%%%%%%%%%%%%%%%%%%%%%%%%%%%%%%%%%%%%%%%%%%%%%%%%%%%%%%%%%%%%%%%%%%%%%%%%%%%%%%%%%%%%%%%%%%%%%%%%%%%%%%%%%%%%%%%%%%%%%%%%%%%%%%%%%%%%%%%%%%%%%%%%%%%%%%%%%%%%%%%%%

\section{Introduction}

It is well known that  Quantum Chromodynamics (QCD) has an approximate U(3)$\otimes$ U(3) chiral symmetry with its subsymmetry U$_A$(1) being explicitly broken by the axial anomaly \cite{Weinberg}. 
In this context, the explicit and spontaneous breaking of chiral symmetry, as well as  the U$_A$(1) anomaly, play a special role, allowing for several nontrivial  assumptions of low energy QCD: (i) the octet of the low-lying pseudoscalar mesons ($\pi,\,K,\,\eta)$ consists of  approximate Goldstone bosons; (ii) the $\eta-\eta^\prime$ phenomenology is characterized by large Okubo-Zweig-Iizuka (OZI) violations. 
In fact, the important contribution of    the  U$_A$(1)-breaking  and the  OZI rule violating terms,  in the process of generation of meson masses and mixing angles, have  been stressed in many phenomenological investigations \cite {Feldmann98,Feldmann02, Ohta80}.
New aspects of mixing and the consistent extraction of mixing parameters from  experimental data have recently  been discussed \cite {Leutwyler}.

It is generally expected that ultra-relativistic heavy-ion experiments will provide the strong interaction conditions which will lead to new physics. In fact, it is believed that the availability of high-energy beams can provides the necessary conditions to observe small-distance scales, allowing to confirm the  QCD as the source of the strong interactions.
Restoration of symmetries and deconfinement  are expected to occur, allowing for the search  of signatures of quark gluon plasma.

The theoretical studies of QCD  at finite temperature and density present    challenging questions, which may be the source of  a productive complement for  understanding relevant features  of particle physics, not only in  heavy-ion collisions, but also in the early universe and in neutron stars.
In particular, the role played by the order of the chiral phase transition on the dynamical evolution of the systems, and possible experimental signs, have recently been  addressed by some authors \cite{scavenius}.
In general, at finite temperature and/or density one expects  chiral symmetry to be restored above a certain temperature (density). 

In QCD, lattice calculations on the nature and order of the phase transitions  indicate that light quarks experience a restoration of chiral symmetry as the temperature increases, with a transition temperature $T_c$ around 150 MeV \cite{lattice1,lattice2,fodor}. 
In the chiral limit, the restoration of  chiral symmetry is signaled by the vanishing of the order  parameters $\left\langle\bar q q\right\rangle$ as the quark masses go to zero. The high temperature phase is sometimes described as a weakly interacting gas of quarks and gluons (plasma phase), which is clearly a simplistic picture for temperatures around  the transition temperature. It has been argued \cite{De Tar} that, just as in the more familiar low temperature phase, the behavior of the high temperature phase is characterized by the propagation of color-singlet objects.  

So far, the more reliable lattice QCD calculations for the phase transition have been focused on the non-zero temperature case. As an alternative to lattice QCD calculations,   QCD-inspired  models have been widely used in recent years to investigate finite temperature and density effects.  

The assumption that the symmetric phase consists of mesonic modes and (deconfined) current quarks underlies the extended version of the  Nambu-Jona-Lasinio (NJL) model \cite{kuni,RKS}. This scenario allows to look for the spectrum of hadrons in parity doubling, whose degeneracy is taken as an indication of an {\em effective} restoration of chiral symmetries. In particular, scalar mesons and its opposite-parity partners, the pseudoscalars, are massive and degenerate in the  symmetric phase. 

In the NJL model we can treat both the scalar and the pseudoscalar mesons on the same  footing. 
The main problem concerning the scalar sector, $J^P=0^+$, which has been under intense investigation over the past few years \cite {Beveren}, is that there are too many light scalars below 1 GeV. The two isoscalars $f_0(600)$ ($\sigma$) and $f_0(980)$  \cite {pdb} as well as the isovector $a_0(980)$ and the isospinor $K_0^*(800)$  (that we will call $\kappa$) \cite {E791} scalars are enough candidates to fill up a nonet of light scalars. 
Although it is accepted that   large 4-quarks and meson-meson components \cite {Close} are necessary to explain  this nonet, here  we shall assume a $q \bar q$ structure for the scalar mesons which are relevant to study the restoration of both chiral and axial symmetries. Recently, Dai and Wu \cite{Day} claimed that ($\sigma,\,f_0,\,a_0,\,\kappa$) can be chiral partners of the pseudoscalar nonet ($\eta,\eta^\prime,\pi,\,K$).
Many other schemes have been suggested to describe the  scalar meson properties. 
In fact, this is a very active field and no definitive conclusion has been reached as to which states are to be considered as $q\bar{q}$, multi-quark, molecule, gluonia or hybrid states \cite {Narison}. 

An important aspect of the problem  is the role played by the anomalously broken U$_A$(1) symmetry in the  restored chiral phase \cite{shuryak,chandrasekharan,Cohen,LeeHatsu,Evans,birse}.  
It has been argued that the chirally restored phase of QCD is effectively symmetric under U(N$_f$)$\otimes$U(N$_f$) rather than SU(N$_f$)$\otimes$SU(N$_f$) at high temperature \cite{Pisarski,shuryak,Evans,Cohen,LeeHatsu}.
Special attention has also been paid to whether or not the effective restoration of the U$_{A}$(1) symmetry and the chiral phase transition occur simultaneously. This question is still controversial and is not settled yet, indicating that we are still far from the full understanding of the dynamics of the processes under discussion. 
Here, we point out two scenarios discussed by Shuryak \cite{shuryak}: in scenario 1, $T_c<<T_{U(1)}$ and  the complete U(N$_f$)$\otimes$U(N$_f$) chiral symmetry is   restored well inside the quark-gluon plasma region; in scenario 2, $T_c\approx T_{U(1)}$. 

The effective restoration of the U$_A$(1) symmetry means that all U$_{A}$(1)-violating effects vanish, i.e., all order parameters of the U$_{A}$(1) symmetry breaking must vanish. 
Since the origin of the anomalous interaction  arises due to  the presence of instantons in the physical state through the 't Hooft term \cite{t Hooft}, the effective restoration of the U$_{A}$(1) symmetry in the NJL model is equivalent to the vanishing of the effects of this interaction.

The question is to look for observables which are strongly influenced by the anomaly and to see if they decrease and eventually vanish, indicating the absence  of the anomaly. One of such quantities is the topological susceptibility, $\chi$, which, in pure color SU(3) theory, can be linked to the $\eta'$ mass trough the Witten-Veneziano formula \cite{Veneziano}. The vanishing of this quantity could be an indication of the restoration of the U$_A$(1) symmetry. In fact, lattice calculations at finite temperature indicate a strong decrease of the topological susceptibility \cite{lattice,latticeChu2}, and recent preliminary results at finite density seems to confirm this tendency \cite{bartolome}.
In addition, since the presence of the axial anomaly causes flavor mixing, with the consequent violation of the OZI rule, both for scalar and pseudoscalar mesons, restoration  of axial symmetry should have relevant consequences for the phenomenology of meson mixing angles, leading to the recovering of the ideal mixing.

In a previous  study \cite{costaUa1}  on effective restoration of chiral and axial symmetries in the NJL model,  we have shown that the axial part of the symmetry is restored before the full U(3)$\otimes$U(3) chiral symmetry. 
Here, we investigate two mechanisms to study an  effective restoration of chiral and axial symmetries, which consists in two different ways for the behavior of the coupling strength of the anomaly.
One of them is based on a phenomenological decreasing \cite{kuni,alkofer}, and the other one is inspired on the behavior of the topological susceptibility as indicated by lattice results at finite temperature \cite{lattice}.
This two cases are going to be compared with two limiting conditions: $g_D={\rm constant}$ and ${ g_D=0}$ from the beginning.
With this methodology we expect to disentangle the competition between U$_A$(1) anomaly and chiral symmetry breaking effects.

After the presentation of the model and the scenarios of restoration of the  axial symmetry in Secs. II and III, respectively, we start our investigation with the study of the consequences of the effective restoration of chiral and axial symmetries with temperature and zero density (Sec. IV).
Due to recent studies on lattice QCD at finite chemical potential it is interesting to investigate also the restoration of the U$_{A}$(1) symmetry at finite density and zero temperature. 
In this case, we will consider two environment scenarios:
completely symmetric matter ($\rho_{u}=\rho_{d}=\rho_{s}$) in Sec. V and  quark matter simulating "neutron" matter in Sec.  VI. Our conclusions are presented in Sec. VII.

%%%%%%%%%%%%%%%%%%%%%%%%%%%%%%%%%%%%%%%%%%%%%%%%%%%%%%%%%%%%%%%%%%%%%%%%%%%%%%%%%%%%%%%%%%%%%%%%%%%%%%%%%%%%%%%%%%%%%%%%%%%%%%%%%%%%%%%%%%%%%%%%%%%%%%%%%%%%%%%%%%

\section{Model and formalism}

We consider the three-flavor NJL type model containing scalar-pseudoscalar
interactions and a determinantal term, the 't Hooft interaction generated by instantons in QCD, which breaks the U$_{A}$(1) symmetry. The model has the following Lagrangian \cite{kuni,RKS}:%

\begin{equation}%
\begin{array}
[c]{rcl}%
\mathcal{L\,} & = & \bar{q}\,(\,i\,{\gamma}^{\mu}\,\partial_{\mu}\,-\,\hat
{m})\,q+\frac{1}{2}\,g_{S}\,\,\sum_{a=0}^{8}\,[\,{(\,\bar{q}\,\lambda
^{a}\,q\,)}^{2}\,\,+\,\,{(\,\bar{q}\,i\,\gamma_{5}\,\lambda^{a}\,q\,)}%
^{2}\,]\\[4pt]
& + & g_{D}\,\,\{\mbox{det}\,[\bar{q}\,(\,1\,+\,\gamma_{5}%
\,)\,q\,]+\mbox{det}\,[\bar{q}\,(\,1\,-\,\gamma_{5}\,)\,q\,]\,\}.
\end{array}
\label{lagr}%
\end{equation}
Here $q=(u,d,s)$ is the quark field with three flavors, $N_{f}=3$, and three
colors, $N_{c}=3$. $\lambda^{a}$ are the Gell-Mann matrices, a =
$0,1,\ldots,8$, ${\lambda^{0}=\sqrt{\frac{2}{3}}\,\mathbf{I}}$. 

Our effective chiral field theory has the same chiral symmetry of QCD, coming out solely from quark interactions. 
The global chiral SU(3)$\otimes$SU(3) symmetry of the underlying Lagrangian (\ref{lagr}) is explicitly   broken  by  the current quark masses $\hat{m}=\mbox{diag}(m_{u},m_{d},m_{s})$.
As the Lagrangian  (\ref{lagr}) defines a non-renormalizable field theory, we introduce a cutoff which sets the 3-momentum scale in the theory.

The NJL model can be generalized to the finite temperature and chemical potential case  by  applying  the Matsubara technique \cite{kapusta} as is shown in the Appendix \ref{apend:Matsubara}.

%%%%%%%%%%%%%%%%%%%%%%%%%%%%%%%%%%%%%%%%%%%%%%%%%%%%%%%%%%%%%%%%%%%%%%%%%%%%%%%%%

\subsection{The gap equation}

In order to put the Lagrangian (1) in a form suitable for bosonization, it is useful to convert the six quark interaction in (\ref{lagr}) into  a four quark interaction \cite{RKS,costaI,costaB,costabig,costaD}, allowing for the  effective quark Lagrangian: 
\begin{eqnarray}
{\cal L}_{eff} &=& \bar q\,(\,i\, {\gamma}^{\mu}\,\partial_\mu\,-\,\hat m)\, q \,\,
\nonumber \\
&+& S_{ab}[\,(\,\bar q\,\lambda^a\, q\,)(\bar q\,\lambda^b\, q\,)]
+\,P_{ab}[(\,\bar q \,i\,\gamma_5\,\lambda^a\, q\,)\,(\,\bar q
\,i\,\gamma_5\,\lambda^b\, q\,)\,],
\label{lagr_eff}
\end{eqnarray}
where  the projectors $S_{ab}\,, P_{ab}$ are presented in the Appendix \ref{apend:formalismo} (Eqs. (\ref{sab}) and (\ref{pab})).

The bosonization  procedure can be done by the integration over the quark fields
in the functional integral with the effective Lagrangian (\ref{lagr_eff}), leading  to  an effective action (\ref{action})  where, as shown in the Appendix \ref{apend:formalismo}, the natural degrees of freedom of low-energy QCD in the mesonic sector are achieved. 
 
The first variation of the effective action leads to the gap equation,
\begin{equation}
M_{i}=m_{i}-2g_{_{S}}\left\langle\bar{q_{i}}q_{i}\right\rangle-2g_{_{D}}\left\langle\bar{q_{j}}q_{j}\right\rangle\left\langle\bar
{q_{k}}q_{k}\right\rangle\,,\label{gap}
\end{equation}
with $i,j,k=u,d,s$ cyclic and $M_{i}$ are the constituent quark masses. The
quark condensates are determined by
\begin{equation}
\left\langle\bar{q}_{i}q_{i}\right\rangle=-i\mbox{Tr}\frac{1}{\hat{p}-M_{i}}=-i\mbox{Tr}\left[
S_{i}(p)\right]\,,
\end{equation}
where $S_{i}(p)$ is the quark Green function. 

%%%%%%%%%%%%%%%%%%%%%%%%%%%%%%%%%%%%%%%%%%%%%%%%%%%%%%%%%%%%%%%%%%%%%%%%%%%%%%%%%

\subsection{Pseudoscalar and scalar meson nonets}

To calculate the meson mass spectrum, we expand the effective action
(\ref{action}) over the meson fields. Keeping the pseudoscalar mesons only, we find the meson masses by using  the rest frame, $\mathbf{P}=0$, and the condition
\begin{equation}
1-P_{ij}\Pi_{ij}^{P}(P_{0}=M,\mathbf{P}=\mathbf{0})=0.\label{rdisp}
\end{equation}
For the non-diagonal mesons $\pi\,,K$, we have
\begin{equation}
P_{\pi}=g_{S}+g_{D}\left\langle\bar{q}_{s}q_{s}\right\rangle, \label{Ppi}%
\end{equation}%
\begin{equation}
P_{K}=g_{S}+g_{D}\left\langle\bar{q}_{u}q_{u}\right\rangle. \label{Pkaon}%
\end{equation}
The polarization operator in Eq. (\ref{rdisp}) takes the form given in the Appendix \ref{apend:formalismo} by Eq. (\ref{ppij}).
The quark-meson coupling  and the meson decay constants $f_{M}$ are also evaluated according to the usual definitions \cite {RKS}.

The inclusion of the 't Hooft interaction in the NJL model allows for flavor mixing, giving rise to a $P^2$-dependent mixing angle $\theta_{P} (P^2)$ \cite {HatK,TNO,shakin,costabig}.
Our scheme for pseudoscalar flavor mixing  consists in the definition of the mixing angle $\theta_P$ in such a way that

\begin{equation}
\left(\begin{array}[c]{r}
\eta\\\eta^\prime
\end{array}
\right)=
O(\theta_P)\left(\begin{array}[c]{r}
\eta_8\\\eta_0
\end{array}
\right)
=\left(
\begin{array}
[c]{rr}%
\mbox{cos}\theta_{P} & -\mbox{sin}\theta_{P}\\
\mbox{sin}\theta_{P} & \mbox{cos}\theta_{P}%
\end{array}
\right) 
\left(\begin{array}[c]{r}
\eta_8\\\eta_0
\end{array}
\right) \,,
\end{equation}
where $\eta$ and $\eta^\prime$ stand for the corresponding physical fields, and $\eta_8$ and $\eta_0$ are the mathematical objects transforming as  octet and singlet states of the SU(3)-flavor pseudoscalar meson nonet, respectively.

The condition to diagonalize $(D_{ab}^{P}(P))^{-1}$ (Eq. (A4)) as ${O}^{-1}(D_{ab} ^{P}(P))^{-1}{O}=\mbox{diag}(D_{\eta}^{-1}(P),D_{\eta^{\prime}}^{-1}(P))$ gives us the equation for the mixing angle:

\begin{equation}
\mbox{tan}\,2\theta_{P}=\frac{2\mathcal{B}}{\mathcal{C}-\mathcal{A}}, \label{anges}
\end{equation}
as well as the inverse meson propagators,

\begin{align}
D_{\eta}^{-1}(P)  &  =\left(  \mathcal{A}+\mathcal{C}\right)  -\sqrt
{(\mathcal{C}-\mathcal{A})^{2}+4\mathcal{B}^{2}}\,,\\
D_{\eta^{\prime}}^{-1}(P)  &  =\left(  \mathcal{A}+\mathcal{C}\right)
+\sqrt{(\mathcal{C}-\mathcal{A})^{2}+4\mathcal{B}^{2}}\,,\label{ipel}
\end{align}
with $\mathcal{A}=P_{88}-\Delta\Pi_{00}(P),\mathcal{C}=P_{00}-\Delta\Pi
_{88}(P),\mathcal{B}=-(P_{08}+\Delta\Pi_{08}(P))$ and $\Delta=P_{00}%
P_{88}-P_{08}^{2}$;
 the different  projectors $P_{ab}$ and  polarization operators $\Pi_{ab}^{P}(P)$ are defined in the Appendix \ref{apend:formalismo} (Eqs. (A11-17)).

In the rest frame,  the condition $D_{\eta}^{-1}(P_{0}=M_{\eta
},\mathbf{P}=0)=0$ and $D_{\eta^{\prime}}^{-1}(P_{0}=M_{\eta^{\prime}}%
,\mathbf{P}=0)=0$ gives as, as usual, the masses for the $\eta$ and $\eta'$.

As shown in other papers, in the framework of the NJL  model \cite{HatK,TNO,shakin,costabig}, since $\cal A, \,B$ and $\cal C$ depend on $P^2$, the mixing angles between the components $\eta_0$ and $\eta_8$, $\theta_P$ (short notation of $\theta_{P} (P^2)$) are $P^2$-dependent.  
In the present paper, when studying temperature and density dependence of several quantities, we  we only discuss the mixing angle for $P^2=M^2_\eta$, for simplicity reasons; we checked that the behavior of the mixing angle for $P^2=M^2_{\eta'}$ gives information qualitively similar.

The same technique used for the  pseudoscalar sector can now be directly applied to the scalar resonances.We deal here with nine scalar resonances: three $a_0$'s, which are the scalar partners of the pions, four $\kappa$'s, being the scalar partners of the kaons, and the $\sigma$ and $f_0$, which are associated similarly with the $\eta$ and $\eta^{\prime}$. As  in the pseudoscalar case,  we have mixing between the $\sigma$ and $f_0$ and the neutral $a_0^{\,0}$. 
Keeping  now the scalar mesons only, we have the effective meson action (\ref{accao2}). The  scalar meson masses  are obtained from the condition

\begin{equation}
1-S_{ij}\Pi_{ij}^{S}(P_{0}=M,\mathbf{P=0})=0,\label{rdispes}
\end{equation}
with%
\begin{equation}
S_{a_0}=g_{S}-g_{D}\left\langle\bar{q}_{s}q_{s}\right\rangle,\label{pci}
\end{equation}
\begin{equation}
S_{\kappa}=g_{S}-g_{D}\left\langle\bar{q}_{u}q_{u}\right\rangle.\label{ppci}
\end{equation}
The polarization operator is presented in the Appendix \ref{apend:formalismo}, Eq. (\ref{ssij}).

Finally we can determine the meson masses of $a_0$ and $\kappa$ using the respective dispersion relations:%
\begin{equation}
1-S_{a_0}{}\Pi_{uu}^{S}(M_{a_0},\mathbf{0})=0,
\end{equation}%
\begin{equation}
1-S_{\kappa}{}\Pi_{us}^{S}(M_{\kappa}{},\mathbf{0})=0.
\end{equation}

For  the diagonal mesons $a_0^{\,0}$, $\sigma$ and $f_0$ we take into account the matrix structure of the propagator in (\ref{accao2}). In the basis of $a_0^{\,0}-\sigma-f_0$ system, we write the projector $S_{ab}$ and the polarization operator $\Pi_{ab}^{S}$ as matrices (see the Appendix \ref{apend:formalismo}).
To find the masses of the $\sigma$ and $f_0$ mesons we use the inverse propagator of the corresponding mesons as indicated in the Appendix \ref{apend:formalismo}. 
The value of the angle $\theta_S$ can also be fixed by a  condition similar to Eq. (\ref{anges}).

When $P_{0}>M_{i}+M_{j}$, i. e., when the mass of the meson exceeds the sum of the masses of its constituent quarks, the meson can decay in its quark--antiquark pairs, being, therefore, a  a resonant state.  Then,  Eqs. (\ref{rdisp}), (\ref{ipel}) and (\ref{rdispes}) have to be calculated in their complex form in order to determine the mass of the resonance $M_{M}$ and the respective decay width $\Gamma_{M}$. Thus, we assume that these set of equations has solutions of the form
\begin{equation}
P_{0}=M_{M}-\frac{1}{2}i\Gamma_{M}, 
\end{equation} 
and, on the other hand, we have to take into account  the imaginary part of the integrals (A8) (for details see Appendix A1).

%%%%%%%%%%%%%%%%%%%%%%%%%%%%%%%%%%%%%%%%%%%%%%%%%%%%%%%%%%%%%%%%%%%%%%%%%%%%%%%%%

\subsection{Vacuum properties and model parameters}

The NJL model exhibits a vacuum phase where chiral symmetry is spontaneously broken, a  mechanism which generates the  constituent quark masses. 
The model is fixed by the coupling constants $g_{S},\,g_{D}$ in the Lagrangian
(\ref{lagr}), the cutoff parameter $\Lambda$ which regularizes momentum space
integrals $I_{1}^{i}$ and $I_{2}^{ij}(P)$, and the current quark masses $m_{i}$.
We start  by considering two sources of chiral  U(3)$\otimes$U(3) symmetry-breaking: (i) current quark masses; and  (ii) U$_A$(1) symmetry-breaking effective interaction.

As already referred, the SU(3) version of the NJL model has five parameters, and we would expect {\em a priori} that one can uniquely fix those parameters in order to fit  five observables $f_\pi$, $M_\pi$, $M_K$, $M_\eta$ and $M_{\eta'}$. However, this is not the case as can be seen comparing the parameter sets of [11] and  [12]. We follow the methodology of Ref. [12] and set $m_u$ to the value 5.5 MeV, and fix the remaining four parameters by fitting   $f_\pi$, $M_\pi$, $M_K$ and $M_{\eta'}$.
The $\eta$ meson in this way is predicted with a mass of 514.8 MeV. This allows for a good overall agreement of our numerical results with the  experimental or phenomenological quantities as shown in Table I.

%%%%%%%%%%%%%%%%%%%%%%%%%%%%%%%%%%%%%%%%%%
\begin{table}[t]
\begin{center}%
\vspace{-0.4cm}
\begin{tabular}
[c]{| c c |}\hline\hline
Explicit symmetry breaking with U$_A$(1) anomaly ($g_D\neq 0$)&\\
\hline\hline
  &  Parameter set \\
Physical quantities & and constituent quark masses\\\hline
$f_{\pi}=92.4$ MeV & $m_{u}=m_{d}=5.5$ MeV\\
$M_{\pi}=135.0$ MeV & $m_{s}=140.7$ MeV\\          
$M_{K}=497.7$ MeV & $\Lambda=602.3$ MeV\\
${M}_{\eta^{\prime}}={960.8}$ MeV & $g_{S}\Lambda^{2}=3.67$\\
${M}_{\eta}=$ ${514.8}$ MeV$^{\ast}$ & $g_{D}\Lambda^{5}=-12.36$\\
$f_{K}=97.7$ MeV$^\ast$ & ${M}_{u}{=M}_{d}=367.7$ MeV$^\ast$\\
${M}_{\sigma}={728.8}$ MeV$^\ast$ & ${M}_{s}=549.5$ MeV$^\ast$\\
${M}_{a_{0}}={873.3}$ MeV$^\ast$ & \\
${M}_{\kappa}={1045.4}$ MeV$^\ast$ & \\
${M}_{f_{0}}={1194.3}$ MeV$^\ast$ & \\
${\theta}_{P}={-5.8}$${{}^{o}}^\ast$\,; ${\theta}_{S}={16}$${{}^{o}}^\ast$ & \\
\hline\hline
\hskip 0.3cm Explicit symmetry breaking without U$_A$(1) anomaly ($g_D= 0$)&\\
\hline\hline
$f_{\pi}=92.4$ MeV & $m_{u}=m_{d}=5.5$ MeV\\
$M_{\pi}=M_{\eta}=135.0$ MeV & $m_{s}=138.75$ MeV\\   
$M_{K}=497.7$ MeV & $\Lambda=602.3$ MeV\\
$f_{K}=95.4$ MeV$^\ast$ & $g_{S}\Lambda^{2}=4.64$\\
${M}_{\eta^{\prime}}={707.5}$ MeV$^\ast$ & $g_{D}\Lambda^{5}=0$\\
${M}_{\sigma}={M}_{a_{0}}={740.1}$ MeV$^\ast$ & ${M}_{u}{=M}_{d}=368$ MeV$^\ast$\\
${M}_{\kappa}={985.38}$ MeV$^\ast$ & ${M}_{s}=587.4$ MeV$^\ast$\\
${M}_{f_{0}}={1194.8}$ MeV$^\ast$ & \\
${\theta}_{P}={-54.74}$${{}^{o}}^\ast$\,;
	${\theta}_{S}={35.264}$${{}^{o}}^\ast$  & \\
\hline\hline
\end{tabular}
\caption{{ Physical quantities in the vacuum state and the  parameter sets for the two symmetry breaking patterns studied in this work. The asterisk signalize predicted  physical quantities.}}
\end{center}
\par
\label{tabelapar}
\end{table}
%%%%%%%%%%%%%%%%%%%%%%%%%%%%%%%%%%%%

However, we point out that this prescription has some problems in which concerns to the description of the $\eta'$ meson. As is well known the NJL model does not confine. Formally, this is reflected by the fact that integrals like  $I(q^2)$, and  hence the polarization function for some mesons, get an imaginary part above the $q \bar q$-threshold that is calculated as indicated at the end of the previous section. 

We will  consider a  second parametrization without U$_A$(1) symmetry-breaking effective interaction (${g_D=0}$) which is also presented in Table I. With this parametrization we also have an overall satisfactory fit to meson properties and quark condensates at zero temperature and density.
However, as expected, the results  show that the anomaly term is necessary to obtain the correct meson mass spectra, especially  by giving the $\eta^\prime$ and $a_0$ its large masses, as well as the  splitting between $\pi$/$\eta$, and  $\sigma$/$a_0$ meson masses. 

%%%%%%%%%%%%%%%%%%%%%%%%%%%%%%%%%%%%%%%%%%%%%%%%%

\section{Scenarios of restoration of the axial symmetry and environment conditions}

Model calculations, for instance within  NJL type or sigma models, generally describe the restoration of  chiral symmetry as a natural consequence of the increase of temperature or density. However, it is found that the observables associated to the anomaly, although decreasing, do not show a tendency to vanish \cite {costaUa1,Bielich,bielich2,roder}. The anomaly in our model is present via the 't Hooft interaction  and its effects appear explicitly  in the gap equations (\ref{gap}) and in the mesons propagators  through products of the anomaly coefficient by  quark condensates (see the expressions of the projectors $S_{a b}$, $P_{ab}$ given by  Eqs. (\ref{sab}) and (\ref{pab})). Such quantities, that act as a kind of "effective anomaly coupling", will be denoted from now on as $\left\langle g_D\right\rangle_i=g_D \left\langle\bar q_i q_i\right\rangle$. The vanishing of such effective coupling should imply the vanishing of the observables associated to the anomaly. However this does not happen  in the present model without being enforced because, while the non-strange quark condensates decrease asymptotically, leading to  an effect almost negligible of $\left\langle g_D\right\rangle_u (\left\langle g_D\right\rangle_d)$, the same does not happen  with $\left\langle g_D\right\rangle_s$, since  restoration of chiral symmetry does  not occur in the strange sector  and $\left\langle\bar q_s q_s\right\rangle$  has always an appreciable value. Therefore, the vanishing of $\left\langle g_D\right\rangle_i$, in general, should be accomplished  by assuming that the anomaly coefficient $g_D$ is a decreasing function of temperature or density.

More attention has been paid, up to now, to the  restoration of axial symmetry with temperature than with density, a motivation which  is supported by the lattice results for the behavior of the topological susceptibility with temperature \cite{lattice}, that indicate a considerable decrease of this quantity. However, the theoretical arguments  concerning the possible restoration  of axial symmetry, whether temperature or density are considered, are similar. While the   difficulties  in testing  the QCD vacuum at high density in heavy-ion collisions are not yet removed, but  expecting that this   will hopefully happen in  future  experiments, it is   useful to have predictions for the non-perturbative regime, even at a qualitative level.  Model calculations in NJL model, although not being an alternative to lattice calculations, can provide a useful contribution. Moreover, lattice calculations  for the behavior of the topological susceptibility with density
\cite{bartolome}, although still in a early stage,  suggest that this  observable is also a decreasing function of density. In view of the considerable interest in the investigation of the behavior of matter at high densities, and the possible restoration of symmetries under these conditions, it is certainly worthwhile to do an exploratory study on  the restoration of the axial symmetry by assuming that  $g_D$ is density dependent, in a  form similar to the temperature dependence.

So, after considering the extreme case of a constant anomaly coupling, $g_D$, we will consider 2 scenarios to study the effective restoration of axial symmetry as summarized in Table II.

%%%%%%%%%%%%%%%%%%%%%%%%%%%%%%%%%%%%%%%%%%
\begin{table}[t]
\begin{center}%
\begin{tabular}
[c]{| c| c c|}\hline\hline
&& Anomaly coefficient $g_D$\\
\hline\hline
{\bf  Case I} & & Constant\\
{\bf Case II} && Fermi function\\
{\bf Case III} && Decreasing exponential\\
\hline\hline
\end{tabular}
\caption{Different schemes of explicit  axial symmetry breaking with temperature (density).}
\end{center}
\par
\label{tabelacas}
\end{table}
%%%%%%%%%%%%%%%%%%%%%%%%%%%%%%%%%%%%%%%%%%

\textbf{Case I:} the anomaly coefficient $g_{D}$ is constant for all range of temperatures or densities. 

\textbf{Case II:} the anomaly coefficient $g_{D}$ is a dropping function of temperature or density. Following the methodology of Ref. \cite{Ohta}, the temperature dependence of $g_{D}$ is extracted by making use of  the lattice results for the topological susceptibility, $\chi$, \cite{lattice}. The expression for $\chi$ in the NJL model is presented in the Appendix \ref{apend:TS}, Eq. (\ref{susc}).  
In view of the arguments presented above,  it seems reasonable to model the density dependence of $g_D$ extrapolating from the results for the finite temperature case  and proceeding by analogy \cite{costaUa1}.

\textbf{Case III:} the anomaly coefficient has the form of a decreasing exponential ($g_{D}(T)=g_{D}(0)$exp$[-(T/T_{0})^{2}]$). This phenomenological pattern of restoration of the axial symmetry  was proposed by Kunihiro
\cite{kuni} in the framework of the present model. 
Here we consider a dependence of the anomalous coupling constant on density  also inspired on the finite temperature scenario. 

We also consider a simplistic scenario without U$_A$(1) anomaly (${\bf g_D=0}$), which is achieved in our model by choosing the second parametrization presented in  Table I. 
We expect that this scenario, being a limiting case, might provide additional information allowing  to understand the interplay between the U$_{ A}$(1) anomaly and flavor symmetry breaking effects.
In fact, in this case the dominant effects come from spontaneous chiral symmetry breaking through quark loop dynamics.

For a more complete understanding of the density effects we will consider two different scenarios  of quark matter: (i) symmetric quark matter; and (ii) neutron matter in $\beta$-equilibrium.
So, the different patterns of axial symmetry with $g_D={\rm constant}$ (Case I), Case II and Case III, and $g_D=0$, are going to be studied in hot media, in symmetric quark matter and in neutron matter.

The restoration of chiral symmetry with temperature or density has been extensively studied in the present model with $g_D$ constant \cite{HatK,RKS,Buballa}. A general conclusion of such studies is that chiral symmetry is effectively restored in the SU(2) sector, but, in the range of densities or temperatures generally considered, the same does not happen in the strange sector. It should be noticed that, as we will show, this conclusion  will not be affected by the different patterns of axial symmetry restoration here considered.

Since in all cases  chiral symmetry is explicitly broken by the presence of non-zero current quark mass terms,  chiral symmetry is realized through parity doubling rather than by massless quarks.
So, the identification of chiral partners and the study of its convergence is the criterion to study the {\em effective} restoration of chiral and axial symmetries. 

%%%%%%%%%%%%%%%%%%%%%%%%%%%%%%%%%%%%%%%%%%%%%%%%%%%%%%%%%%%%%%%%%%%%%%%%%%%%%%%%%%%%%%%%%%%%%%%%%%%%%%%%%%%%%%%%%%%%%%%%%%%%%%%%%%%%%%%%%%%%%%%%%%%%%%%%%%%%%%%%%%

\section{Results for the mesonic behavior at finite temperature and zero density}

%%%%%%%%%%%%%%%%%%%%%%%%%%%%%%%%%%%%
\begin{figure}[t]
\begin{center}
\begin{tabular}
[c]{cc}%
\includegraphics[width=0.53\textwidth]{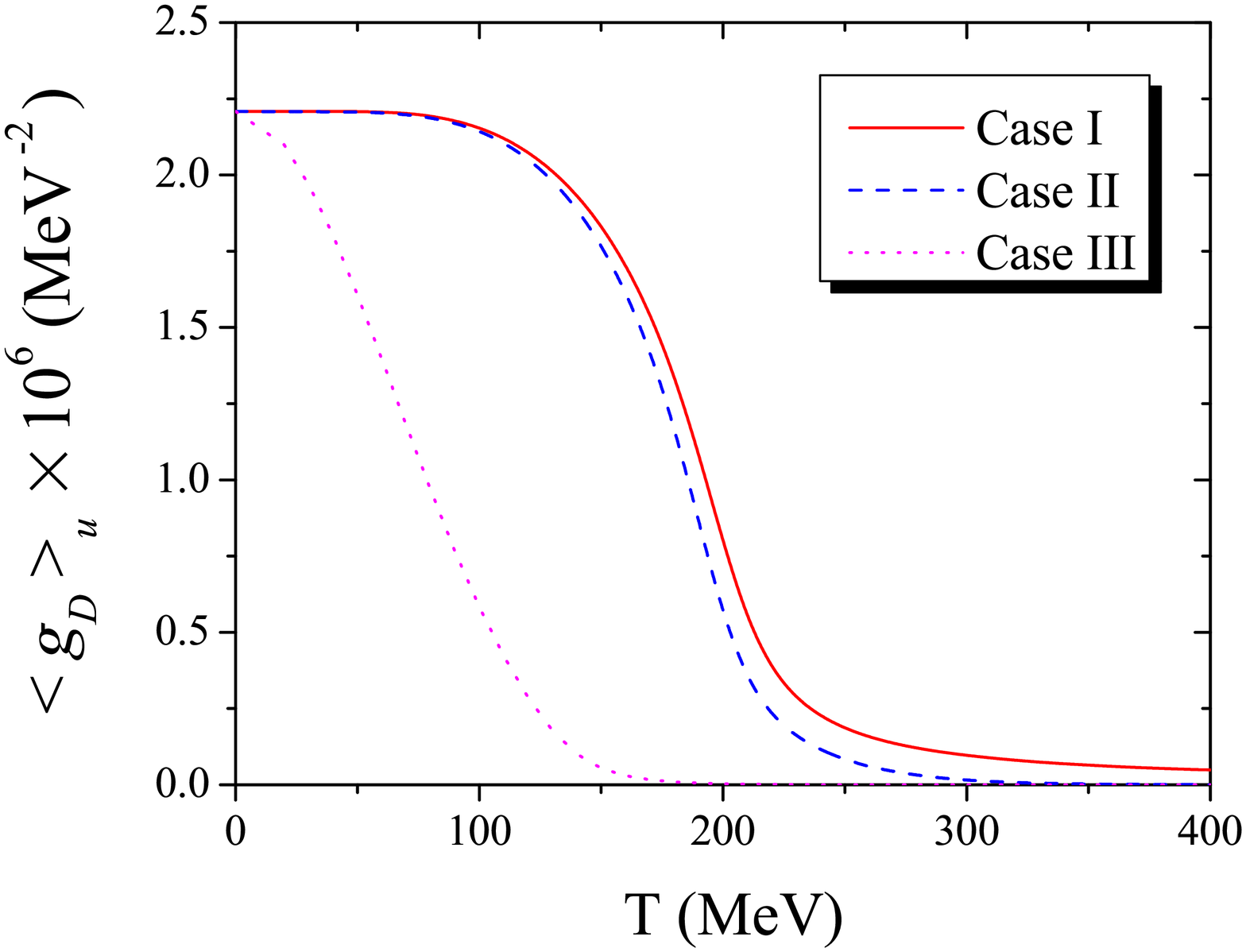} &
\hspace{-1.0cm}\includegraphics[width=0.53\textwidth]{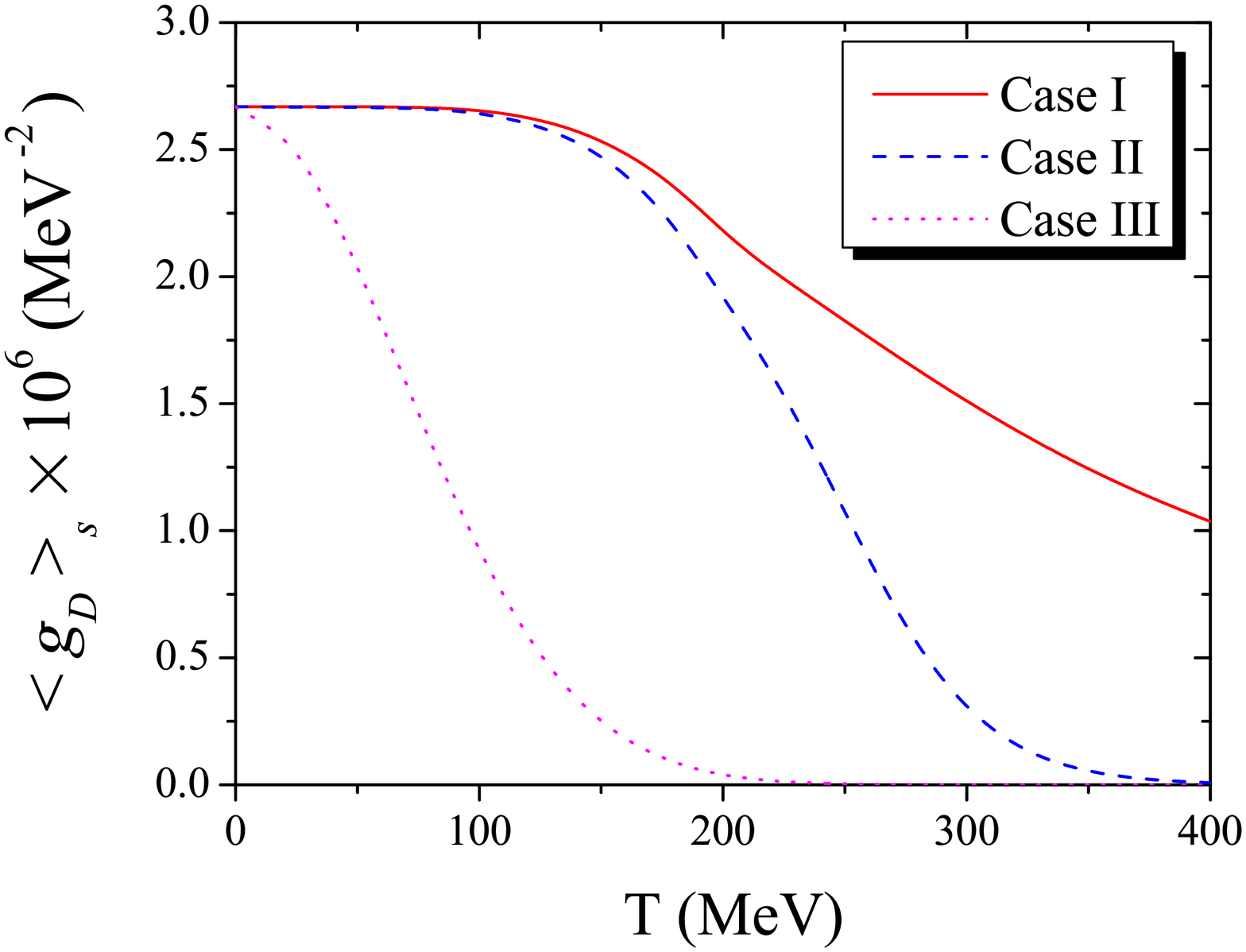}
\end{tabular}
\end{center}
\caption{Temperature dependence of $\left\langle g_D\right\rangle_u $ (left panel) and $\left\langle g_D\right\rangle_s $ (right panel) for the different cases. }%
\label{gDTemp}%
\end{figure}
%%%%%%%%%%%%%%%%%%%%%%%%%%%%%%%%%%%%

In this section we analyze  the mesonic behavior at finite temperature and zero chemical potentials. 
A significant feature of this analysis is that the observables, that depend on the anomaly coupling only via $\left\langle g_D\right\rangle_{u}$ ($\left\langle g_D\right\rangle_{d}$), are not significantly affected by the  specific temperature  dependence of $g_D$, in the high temperature  region,  because chiral symmetry is approximately  restored with the consequent asymptotic vanishing of the non-strange quark condensates. In order to see the importance of the behavior of the effective anomaly coupling for the quantities under study, we plot them in Fig. \ref {gDTemp}. 
   
%%%%%%%%%%%%%%%%%%%%%%%%%%%%%%%%%%%%%%%%%%%%%%%%%%%%%%%%%%%%%%%%%%%%%%%%%%%%%%%%%

\subsection{Explicit chiral symmetry breaking with U$_{A}$(1) anomaly}

%%%%%%%%%%%%%%%%%%%%%%%%%%%%%%%%%%%%
\begin{figure}[t]
\begin{center}
\includegraphics[width=0.73\textwidth]{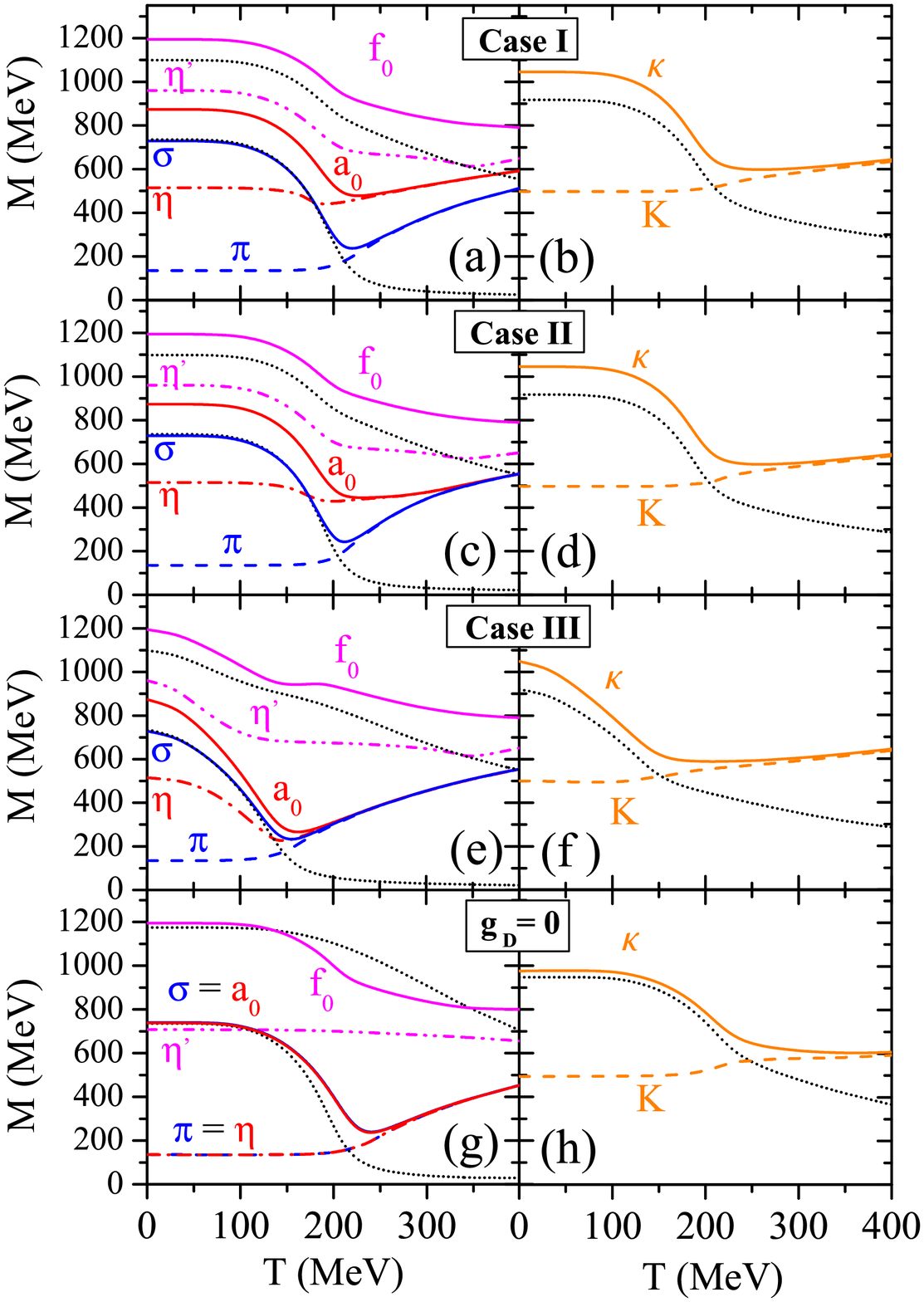}
\end{center}
\caption{Temperature dependence of meson masses and of the continuum thresholds (dotted lines) $2M_u,\,2M_s$  (left panels) and $M_u+M_s$.}%
\label{tempmass}%
\end{figure}
%%%%%%%%%%%%%%%%%%%%%%%%%%%%%%%%%%%%

\textbf{Case I.} We will start with  {Case I} (see Fig. \ref{tempmass}(a) and (b)) that will be compared with the other cases under discussion.
In the panel (b) we have the meson masses for the $K$-meson and its chiral
partner  $\kappa$. In the panel (a) we plot the other scalar and pseudoscalar mesons: ($\sigma, \,f_{0},\,a_{0}, \,\pi, \,\eta, \,\eta^{\prime}$). In both panels, and for all graphics, the dotted line means the respective continuum. The crossing of the $\pi$ and $\eta$ lines with the quark threshold $2M_{u}$, and the $K$ line with $M_{u}+M_{s}$ indicates the respective Mott transition temperature, $T_{M}$. Mott transition comes from the fact that mesons are not elementary objects but are composed states of $q\bar q$ excitations, and is defined by the transition from a bound state to a resonance in the continuum of unbound states. Above the Mott temperature we have taken into account the imaginary parts of the integrals $I_{2}^{ij}$ and used a finite width approximation \cite{RKS,costabig}.

\begin{figure}[t]
\begin{center}
\begin{tabular}
[c]{cc}%
\includegraphics[width=0.53\textwidth]{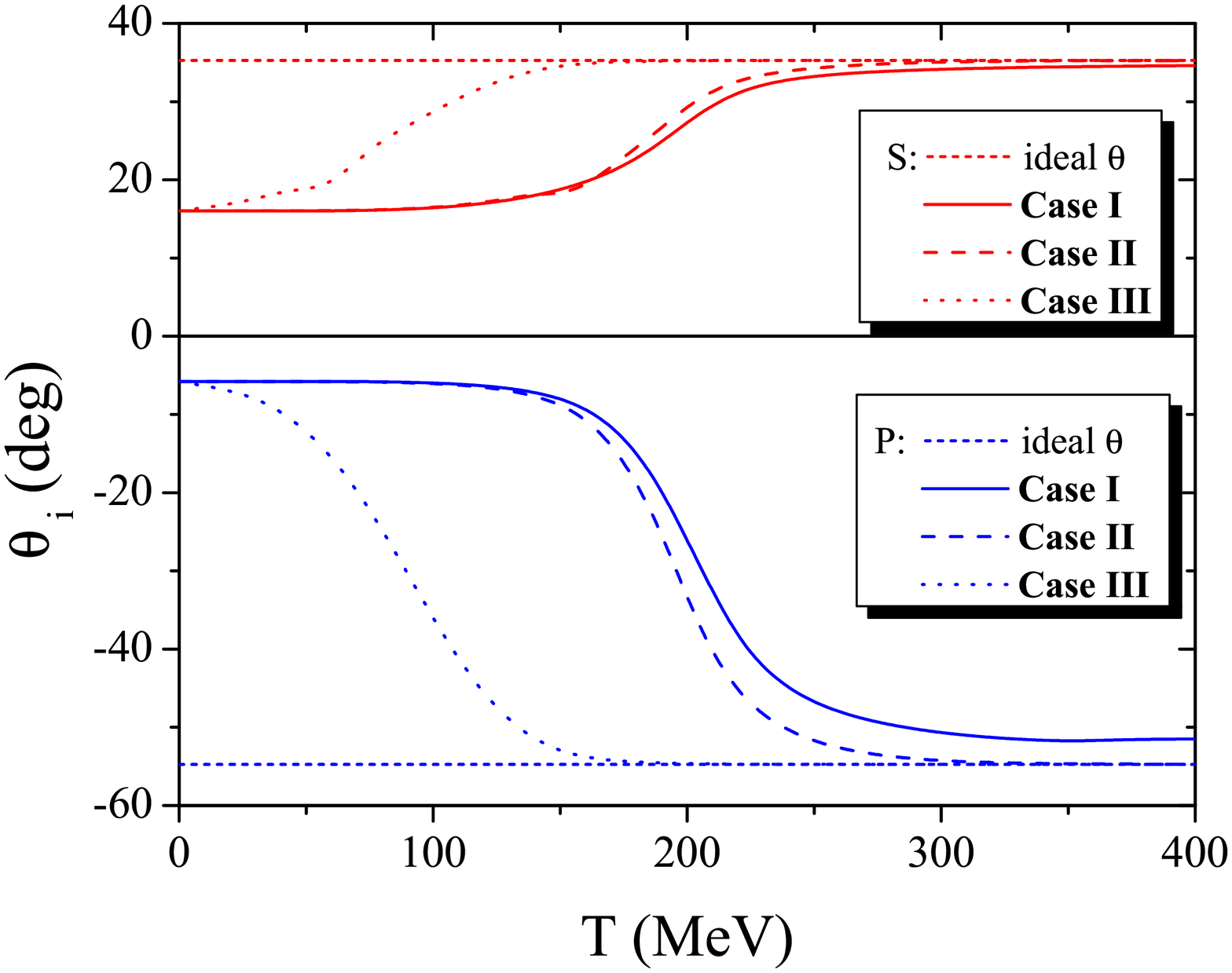} &
\hspace{-1.0cm}\includegraphics[width=0.54\textwidth]{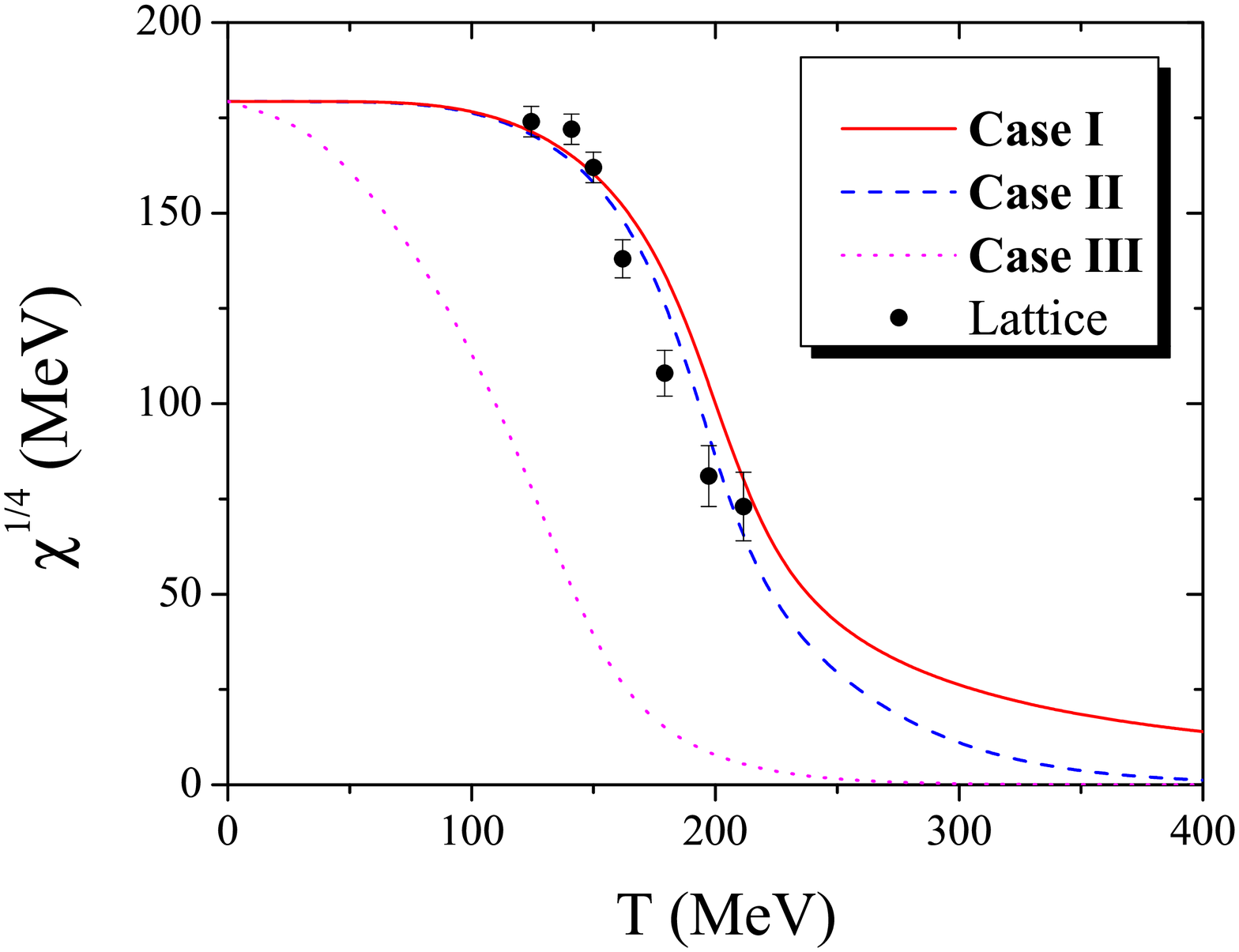}
\end{tabular}
\end{center}
\caption{Left panel: scalar and pseudoscalar mixing angles as a function of temperature for  the three cases  presented in Table II, and for the ideal mixing. Right panel: Topological susceptibility as a function of temperature for the three cases. The  lattice data results are plotted with error bars \cite{lattice}.}%
\label{tempang}%
\end{figure}

Let us summarize here the behavior of the pseudoscalar mesons and analyze what this behavior can tell us about possible restoration of symmetries. One can see that Mott temperatures for $\eta$ and $\pi$ mesons are: $T_{M_\eta}=180$ MeV and $T_{M_\pi}=212$ MeV. The $\pi$ and $K$ mesons become unbound at approximately the same temperature: $T_{M_K}=210$ MeV. On the  other side, the $\eta^{\prime}$ is always above the continuum $\omega_{u}=2M_{u}$, and $\eta$ has always a strange component for all temperatures, once its mixing angle $\theta_{P}$ never gets  the ideal  value: $\theta_{P}=$ $-54.736^{\circ}$ (see Fig. \ref{tempang}).

%%%%%%%%%%%%%%%%%%%%%%%%%%%%%%%%%%%%
%%%%%%%%%%%%%%%%%%%%%%%%%%%%%%%%%%%%

Concerning the scalar sector, we notice that the $\sigma$ meson is the only scalar meson that is a bound state for small temperatures (the others are always resonances) but turns into a resonance at $T_{M_{\sigma}}\simeq 160$ MeV. This meson has a strange component that decreases with temperature but never vanishes since the ideal mixing angle, $\theta_s=35.264^{\circ}$, is never attained in the range of temperatures studied (see Fig. \ref{tempang})

For $T\gtrsim250$ MeV the $\sigma$ starts to be degenerate with the $\pi$.
As for the $a_{0}$ meson,  it is always a non-strange state and is  above the continuum $\omega_{u}=2M_{u}$. It can be seen in Fig. \ref{tempmass}(a) that the partners ($\pi,\sigma$) and ($\eta,a_{0}$) become degenerate  at almost the same temperature. In both cases, this behavior is the signal  of the effective restoration of chiral symmetry in the non-strange sector.
Distinctly, the $\eta^{\prime}$ and $f_{0}$ masses do not show a tendency to
converge in the region of temperatures studied. We interpret this behavior as an indication that chiral symmetry does not show tendency to get restored in the  strange sector (see $2M_s$, upper dotted curve in Fig. \ref{tempmass}(a)).

Finally, we focus on   the $\kappa$-meson (Fig. \ref{tempmass}(b)). It is always an unbound state and, as the temperature increases, it shows  tendency to get degenerate in mass with the $K$-meson. 
For comparison purposes, we summarize in Table III an overview of the transition temperatures of the effective restoration of chiral (second line) and axial (third line) symmetries in the different cases studied in the present paper. The masses of the corresponding chiral partners become degenerate above the referred temperatures. 

Summarizing, the SU(2) chiral partners ($\pi,\sigma$) and ($\eta,a_0$) become degenerate at $T\simeq 250$ MeV; the chiral partner ($K,\kappa$) converges at $T\simeq 350$ MeV and  ($\eta^\prime, \,f_0$) do not show a tendency to converge in the region of temperatures studied.

As expected, the axial symmetry is not restored at high temperatures and the topological susceptibility is also far away from being  zero (see Fig. \ref{tempang}).

%%%%%%%%%%%%%%%%%%%%%%%%%%%%%%%%%%%%
\begin{table}[t]
\begin{center}%
\begin{tabular}
[c]{| c| c| c| c| c|}\hline\hline
& Case I & Case II & Case III & $g_D=0$\\
& ($g_D={\rm constant}$) &&& \\
\hline\hline
 SU(2) chiral-    & 250 MeV & 250 MeV & 225 MeV  & 300 MeV\\
 transition temperature &&&&\\
 \hline
 U(2) axial-&---& 350 MeV & 225 MeV & ---\\
 transition temperature &&&&\\
\hline\hline
\end{tabular}
\caption{ Transition temperatures of the effective restoration of chiral  and axial  symmetries in the different cases.}
\end{center}
\par
\label{tabelatemp}
\end{table}
%%%%%%%%%%%%%%%%%%%%%%%%%%%%%%%%%%%%

\textbf{Case II.} Some of the results for Case II have been presented in \cite{costaUa1}. Here we summarize the conclusions obtained. 

As $m_{u}=m_{d}<<m_{s}$, the (sub)group SU(2)$\otimes$SU(2) is a much better symmetry of the Lagrangian (1) than SU(3)$\otimes$SU(3). So, the effective restoration of the SU(2) symmetry implies the degeneracy between the chiral partners $(\pi,\sigma)$ and $(\eta,a_{0})$ which is verified around $T\simeq 250$ MeV (see Fig. \ref{tempmass}(c)  and Table III). For temperatures at $T\simeq 350$ MeV both $a_{0}$ and $\sigma$ mesons become degenerate with the $\pi$ and $\eta$ mesons, showing, as explained below, an effective restoration of both chiral and axial symmetries. Without the restoration of U$_{A}$(1) symmetry ({Case I}), the $a_{0}$ mass was moved upwards and never met the $\pi$ mass, the same argument being valid for the $\sigma$ and $\eta$ mesons. We remember that the determinant term acts in an opposite way for the scalar and pseudoscalar mesons as can be seen, for instance, in Eqs. (\ref {Ppi})  and  (\ref {pci}). So, only after the effective restoration of the U$_{A}$(1) symmetry we can recover the SU(3) chiral partners $(\pi,a_{0})$ and $(\eta,\sigma)$ which are now all degenerate. 
This is compatible with scenario 1 of Shuryak \cite{shuryak}: the signals for the effective restoration of the axial symmetry occur at a temperature where the signals of the full restoration of U(3)$\otimes$U(3) symmetry are not yet visible.  
In fact, the $\eta^{\prime}$ and $f_{0}$ masses do not  show a clear tendency to converge in the region of temperatures studied, this absence of convergence being probably due to the fact that, in the region of temperatures above $T\simeq 350$ MeV, those mesons are purely strange and the chiral symmetry in the strange sector is far from being effectively restored.

The analysis of the temperature dependence of the mixing angles in Fig. \ref{tempang}, allowing for a better understanding of the meson behavior through the evolution of the quarkonia content, provides further indication of the restoration of the axial symmetry: $\theta_{S}$ ($\theta_{P}$) starts at $16^{\circ}$ ($-5.8^{\circ}$) and goes, smoothly, to the ideal mixing angle $35.264^{\circ}$ ($-54.74^{\circ}$).
This means that flavor mixing no more exists. 
In fact,  referring to  the SU(2) chiral partners ($\pi,\sigma$) and ($\eta,a_0$), we found that the $a_{0}$ and  $\pi$ mesons are always purely non-strange quark systems, while the $\sigma$ ($\eta$) meson becomes purely non-strange when $\theta_{S}$ ($\theta_{P}$) goes to $35.264^{\circ}$ ($-54.74^{\circ}$), at $T\simeq 350$ MeV. 

Analyzing the ($K$,$\kappa$) chiral partner, we conclude that the behavior of the  mesons is not  significantly influenced by the type of temperature dependence of $g_{D}$ used here, as expected. In fact, in the range of temperatures where $g_{D}(T)$ could be important,  $M_{s}$ does not change appreciably, and we know that these meson masses are very sensitive to $M_{s}$. For the range of temperatures where the $(\pi,a_{0})$ and $(\eta,\sigma)$ chiral partners become degenerate, the strange quark mass of  $M_{s}$ is already independent of the $g_{D}$ dependence of the temperature (see $2M_s$, upper dotted curve in Fig. \ref{tempmass}(c)).
This is due to the fact that, as explained before, $M_s$ depends on the anomaly through $\left\langle g_D\right\rangle_u$. 

We notice that our analysis of the effective restoration of symmetries is
based on the degeneracy of chiral partners that occurs in a region of temperatures where the mesons are no more bound states (they dissociate in
$q\bar{q}$ pairs at their respective Mott temperatures \cite{RKS,costabig}).
Moreover, the mesons $\eta^{\prime}$ and $f_{0}$ are $q\bar{q}$ resonances
from the beginning and its description is unsatisfactory.

Summarizing, we conclude that at $T\simeq 250$ MeV the SU(2) chiral partners become degenerate in mass, whereas at  $T\simeq 350$ MeV, the same happens with ($\pi,\, \sigma,\, \eta,\, a_{0}$) mesons: the OZI rule is restored and $\chi$ goes asymptotically to zero (Fig. \ref{tempang}, dashed line of right panel). These results indicate an effective restoration of the U$_{A}$(1) symmetry.

\textbf{Case III.} Finally, we analyze {Case III } that  is  similar to {Case II} as we can see in Fig. \ref{tempmass}. The main difference is that the temperature dependence of $g_{D}$ used does strengthen significantly the chiral phase transition.
In fact, the SU(2) chiral partners $(\pi,\sigma)$ and $(\eta, a_{0})$ are all degenerate for $T\simeq 225$ MeV ($T\simeq 250$ MeV in Case II). Linking this fact to the behavior of the $\chi$ (in Fig. \ref{tempang}, dotted line in right panel) that  goes very fast  to zero, being zero at about $250$ MeV, and with the behavior of the mixing angles (Fig. \ref{tempang}, dotted lines in left panel), $\theta_{P}$ and $\theta_{S}$, that go both to its ideal values at $200$ MeV, we conclude that both symmetry restorations happen around the same temperature.

The comparison between Case I ($g_D={\rm constant}$) and Case III helps to understand this situation. We observe that the more rapid decrease of the temperature dependence of $M_s$ in Case III, till $T\approx 250$ MeV, indicated by the upper dotted lines ($2 M_s$) in the left panel of Fig. \ref{tempmass} (see Fig. \ref{tempmass}(a), (e)), cooperates with the decreasing of $g_D(T)$ allowing for the restoration of chiral and axial symmetries at the same temperature $T\simeq 225$ MeV.
This can also be seen, for instance, in the behavior of the effective anomaly coupling $\left\langle g_D\right\rangle_s$ that goes to zero at almost the same temperature (see Fig. \ref{gDTemp}).
This is in accordance with scenario 2 of Shuryak \cite{shuryak}.
The existence of  cooperative effects of restoration of chiral and axial symmetries has  already been noticed  by Kunihiro \cite{kuni}, who report a situation where the axial symmetry is restored before chiral symmetry, a  scenario usually considered not realistic \cite{shuryak}.

Concerning ($K$,$\kappa$) chiral partner, the conclusions are similar to those of Case II; the only difference is a faster decrease of the splitting in the low temperature region, due to the faster decrease of $M_s$.

%%%%%%%%%%%%%%%%%%%%%%%%%%%%%%%%%%%%%%%%%%%%%%%%%%%%%%%%%%%%%%%%%%%%%%%%%%%%%%%%%

\subsection{Explicit chiral symmetry breaking without U$_{A}$(1) anomaly}

We consider now a simplistic scenario without U$_A$(1) anomaly (${\bf g_D=0}$), which is achieved in our model by choosing the second parametrization presented in  Table I. 

We start with the $\pi$-meson that, as expected, is always degenerate with $\eta$.
In fact, the $\eta $-meson is a pure non-strange state for all temperatures, with a ideal mixing angle $\theta_{P}=$ $-54.736^{0}$.

On the  other side, the $\eta^{\prime}$ is always a pure strange state which crosses the continuum $\omega_{u}=2M_{u}$  for $T\gtrsim110$ MeV, becoming then a resonance state. Like  in the $g_D\neq 0$ cases, the $\eta^{\prime}$ meson shows no tendency to become degenerate with $f_{0}$, a consequence of the insufficient restoration of chiral symmetry   in the strange  sector, as it has already been noticed.
The $a_{0}\,(\equiv \sigma)$ is always a non-strange state, it is always above the continuum $\omega_{u}=2M_{u}$. 

As the temperature increases, due to the absence of the U$_A$(1) anomaly, the members of the chiral pairs ($\pi,\sigma$) and ($\eta,a_{0}$)  become all degenerate simultaneously ($T\simeq 300$ MeV), reflecting the effective 
restoration of chiral symmetry in the non-strange sector. 
We notice that, as indicated in Table III, this is the case where the transition temperature to the  SU(2)$\otimes$SU(2) symmetry is higher, indicating that, as already referred, the anomalous coupling can be important to drive  the effective restoration of the chiral symmetry itself.  

Concerning the kaon and its chiral partner $\kappa$, they show a clear tendency
to get degenerate, but at temperatures that are higher than in the previous cases. 

Summarizing, the high temperature regime ($T\approx 300$ MeV)  in {Cases II} and III, where the axial symmetry is effectively restored,  and the situation ${ g_D=0}$ are very similar: the SU(3) chiral partners ($\pi,a_0$) and ($\eta,\sigma$) are degenerate, and the $\eta^\prime$ and $f_0$ mesons have similar splittings. The more relevant differences in the behavior at lower temperatures are manifestations of the different role played by the axial anomaly and the dynamical flavor symmetry breaking effects. For instance, the constituent strange quark mass has a very different behavior in the three scenarios as can be seen by the curve representing $2\,M_s$ (upper dotted curve) in Fig. \ref{tempmass} ((c), (e) and (g)).
In addition, we notice that, differently from Case II, in Case III the restoration the U$_A$(1) symmetry drives of chiral symmetry. 

%%%%%%%%%%%%%%%%%%%%%%%%%%%%%%%%%%%%%%%%%%%%%%%%%

\section{Results for symmetric quark matter}

In order to study the effective restoration of chiral and axial symmetries at finite density, we start by considering a completely symmetric quark matter ($\rho_{u}=\rho_{d}=\rho_{s}$).
Before our analysis, let us make some considerations about this type of matter.
Although rather schematic, this case simulates a situation where the hypothesis of absolutely stable strange quark matter (SQM) can be explored \cite{costabig}. It has been argued
\cite{Buballa,Faessler} that SQM may only be stable if it has a large fraction
of strange quarks ($\rho_{s}\approx \rho_{u}\approx \rho_{d}$). The speculations
on the stability of SQM are supported by the  observation that the inclusion of the strangeness degree of freedom allows for a larger decrease of the strange quark mass which can produce a sizable binding energy. In \cite{costabig} we have confirmed this tendency when compared with neutron matter.
We notice  that there are always strange valence quarks present, so the strange quark mass decreases more strongly, although, even in this case, it is still away from the strange current quark mass  for high densities \cite{costabig}.
The advantage of considering  this type of matter is that, like in the non-zero temperature case,  all three pions and all four kaons are degenerate in medium (contrarily to what happens for neutron matter in $\beta$-equilibrium  to be discussed in the next section).
So, the present environment  can  provide fruitful comparisons with the non-zero temperature case.

%%%%%%%%%%%%%%%%%%%%%%%%%%%%%%%%%%%%%%%%%%%%%%%%%%%%%%%%%%%%%%%%%%%%%%%%%%%%%%%%%

\subsection{Explicit chiral symmetry breaking with U$_{A}$(1) anomaly}

%%%%%%%%%%%%%%%%%%%%%%%%%%%%%%%%%%%%
\begin{figure}[t]
\begin{center}
\includegraphics[width=0.73\textwidth]{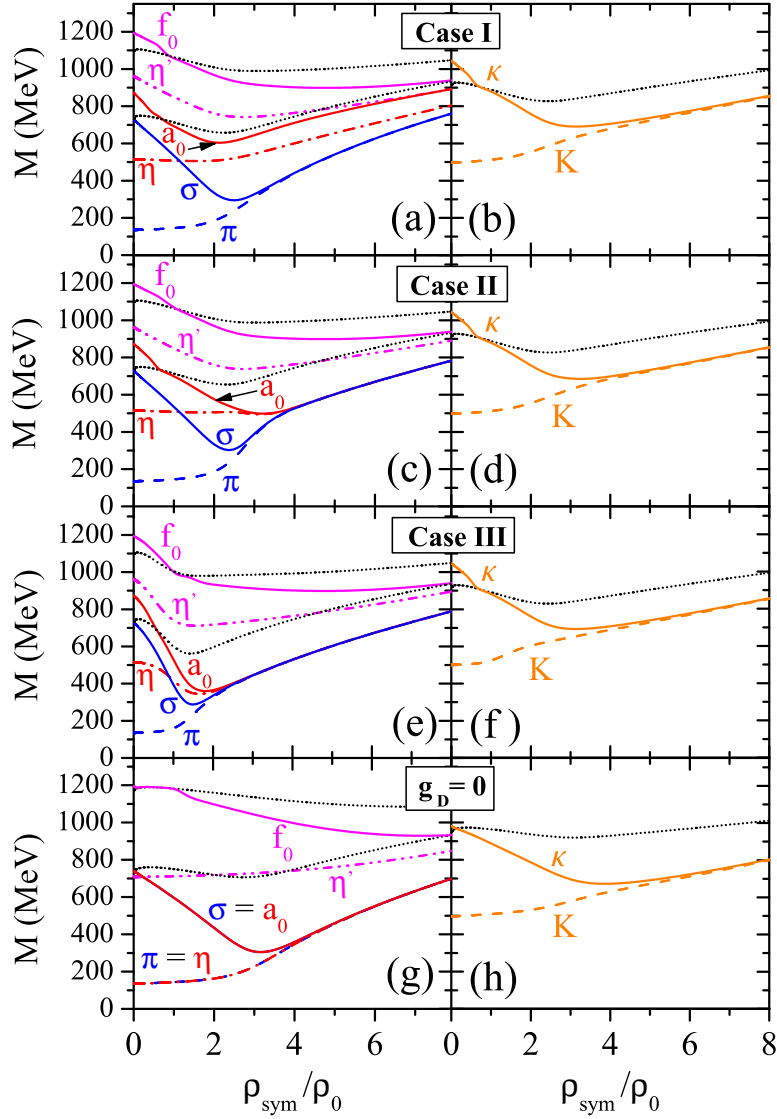}
\end{center}
\caption{{Density dependence of meson masses and of  limits of the Dirac sea continua (dotted lines) defining $q\bar q$ thresholds for the mesons $\eta^\prime,\,a_0,\,\kappa$. The $f_0$ meson is always a resonance state.}}%
\label{symmass}%
\end{figure}
%%%%%%%%%%%%%%%%%%%%%%%%%%%%%%%%%%%%

The study of density effects is performed using a methodology  analogous to the temperature case. 
So we will also consider  3 scenarios for the behavior of the anomalous coupling $g_D$  (see Table II).
Some conclusions are very similar to the temperature case as we can check in Fig.
\ref{symmass}, so we will concentrate on the main differences. 

\textbf{Case I.} 
In this case  we have an interesting phenomenon:  the $a_{0}$  does not degenerate with the $\eta$-meson, but with the $\eta^\prime$-meson as we can see in Fig. \ref{symmass}(a). In fact, the $\eta^{\prime}$-meson, that starts as an unbound state and becomes bound for $\rho_{sym}\gtrsim4.5\rho_{0}$, degenerates with the $a_{0}$-meson for higher densities. This is due to the presence of strange valence quarks in the medium, which causes $M_{s}$  to decrease more strongly \cite{costabig}, so the influence of the $s$ sector will be lower in the mass of the $\eta^{\prime}.$ 

\begin{figure}[t]
\begin{center}
\begin{tabular}
[c]{cc}%
\includegraphics[width=0.53\textwidth]{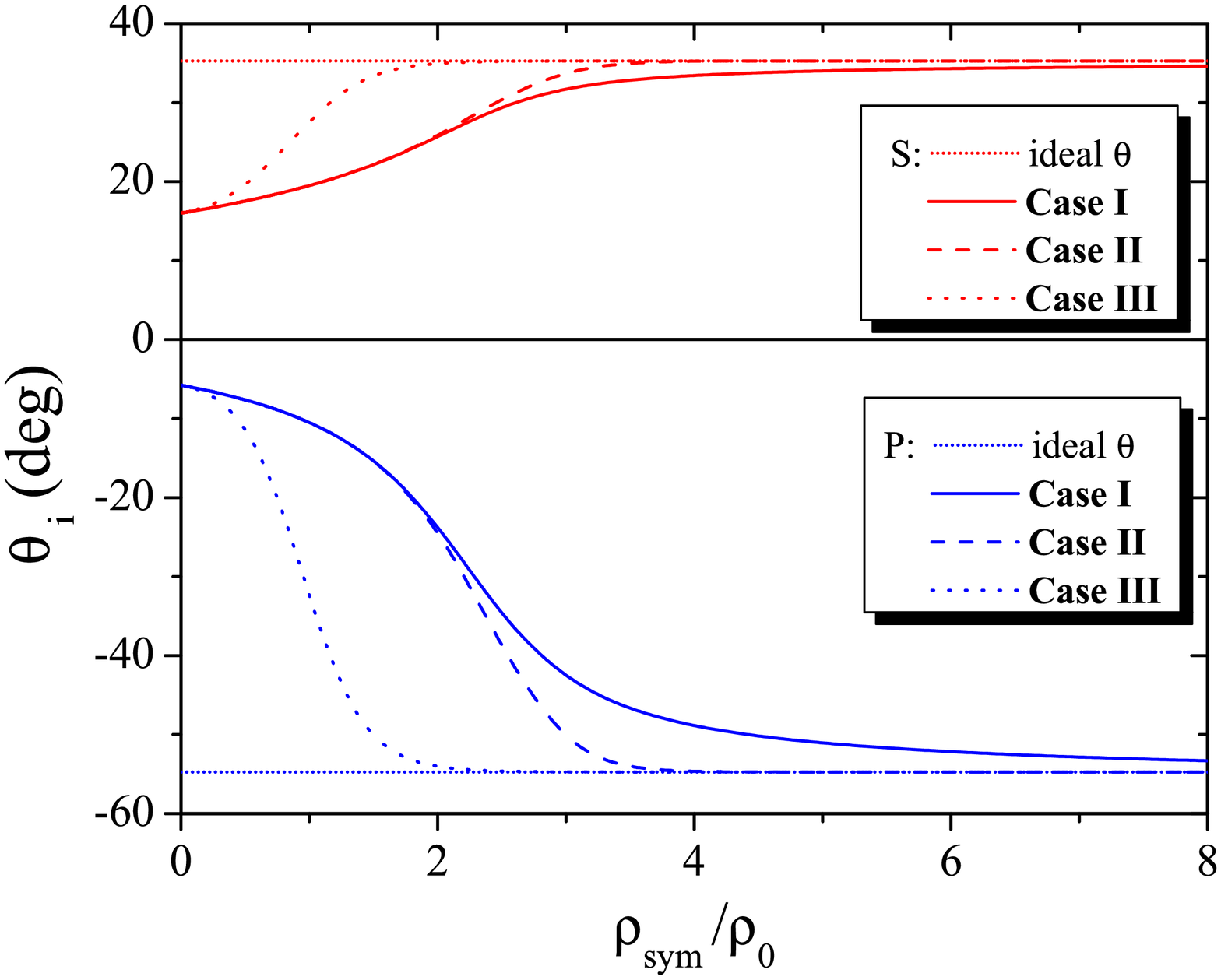} &
\hspace{-1.0cm}\includegraphics[width=0.54\textwidth]{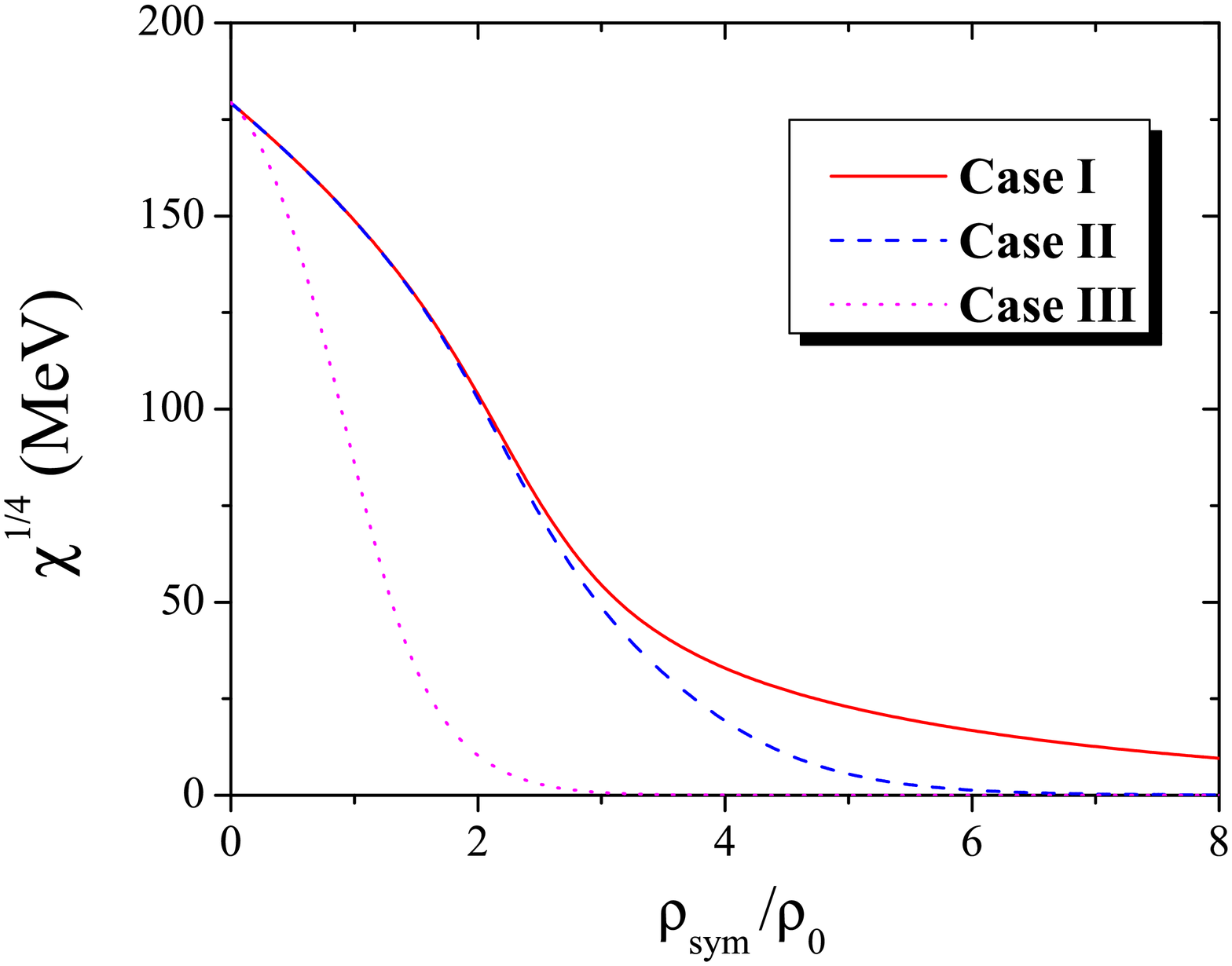}
\end{tabular}
\end{center}
\caption{Left panel: scalar and pseudoscalar mixing angles as a function of density for   the three cases  presented in Table II, and for the ideal mixing ($g_D=0$). Right panel: topological susceptibility as a function of density for the three cases.}%
\label{symang}%
\end{figure}

\textbf{Case II.} The study of {Case II} in symmetric quark matter is inspired, as already referred, in the previous Case II at finite temperature. 
So, we postulate a dependence for $\chi$ formally similar to the temperature case as is shown in Fig. \ref{symang} (dashed line), i.e., using a Fermi function. With this topological susceptibility we obtain the density dependent anomalous coupling $g_{D}(\rho_{sym})$.

Using this density dependence  we arrive at conclusions similar to the finite temperature case discussed before. The chiral partners $(\pi\,,\sigma)$ become degenerate at a density $\rho_{sym}\simeq 3.5 \rho_0$ and the same happens to the chiral partners $(\eta\,,a_0)$ (Fig. \ref{symmass}(c)); this density is the onset for effective restoration of chiral symmetry in the  SU(2) sector.

The analysis of the mixing angles (Fig. \ref{symang}) indicate that at $\rho_{sym}\simeq 4\rho_0$ the scalar and pseudoscalar mixing angles reach its ideal values and, consequently, the $\eta$ and the $\sigma$ become purely non strange. At this density the $\eta'$ becomes purely non strange and does not show a tendency to degenerate with $f_0$,  as in the finite temperature case. 

Summarizing, as the density increases, the chiral partners ($\pi,\sigma$) and ($\eta,a_{0}$) become degenerate (for $\rho_{sym}\simeq 4\rho_{0})$. Associating this with the behavior of the $\theta_{P}$ and $\theta_{S}$ mixing angles and the behavior of the chiral susceptibility, that goes to zero (dashed line in Fig. \ref{symang}), we conclude that we have an effective restoration of U$_{A}$(1) symmetry in this situation.

%%%%%%%%%%%%%%%%%%%%%%%%%%%%%%%%%%%%%%%%%%
%%%%%%%%%%%%%%%%%%%%%%%%%%%%%%%%%%%%

\textbf{Case III.} Similarly, for {Case III} we postulate the following dependence for $g_{D}:$ $g_{D}(\rho_{sym})=g_{D}(0)$exp$[-(\rho_{sym}/\rho_{0})^{2}],\ $ which is inspired in the corresponding finite temperature scenario.
The topological susceptibility with this dependence of the coupling anomaly is plotted in Fig. \ref{symang}, dotted line.
This case is very similar to  {Case II}  and the overall conclusions are parallel to  the finite temperature case (to compare see Fig. \ref{tempmass}(e), (f)). The density dependence for $g_{D}(\rho_{sym})$ that we used also does strengthen the chiral phase transition: both symmetry restorations happen simultaneously for slightly lower densities ($\rho_{sym}\simeq 3.0\rho_{0}$).

In fact, the SU(3) chiral partners $(\pi,a_{0})$ and $(\eta,\sigma)$ are all degenerate at very earlier values of the density ($\rho_{sym}\simeq 3.0\rho_{0}$), compared with {Case II} ($\rho_{sym}\simeq 4.0\rho_{0}$). This results from the behavior of  $\chi$ (in Fig. \ref{symang}, dotted line in right panel) that  goes to zero for $\rho_{sym}\simeq 3\rho_{0}$, and by the behavior of the mixing angles (Fig. \ref{symang}, dotted line in left panel) where both, $\theta_{P}$ and $\theta_{S}$, go to the ideal mixing angles for $\rho_{sym}\simeq 2.5\rho_{0}$.

Concerning  the ($K$,$\kappa$) partners (Fig. \ref{symmass}(f)), we conclude that in all three cases their behavior is very similar: they practically do not  depend on the shape of $g_{D}$.

%%%%%%%%%%%%%%%%%%%%%%%%%%%%%%%%%%%%%%%%%%%%%%%%%%%%%%%%%%%%%%%%%%%%%%%%%%%%%%%%%

\subsection{Explicit chiral symmetry breaking without U$_{A}$(1) anomaly}

Finally, we analyze  the behavior of mesonic modes in the absence of the axial anomaly: ${\bf g_{D}=0}$.
Looking at the  $\pi$-meson  behavior plotted  in  Fig. \ref{symmass} (g), we conclude that the pion  is always degenerate with $\eta$ and they are always bound states. The $\eta$ ($\eta^\prime$) meson is a pure non-strange (strange)  state for all range of densities. For $2\rho_0 \lesssim\rho_{sym}\lesssim 4\,\rho_0$ the $\eta^{\prime}$-meson  is a resonance state as can be seen in Fig. \ref{symmass}(g).
As in the {Cases I}, {II} and {III}, the $\eta^{\prime}$-meson does not show  tendency to become degenerate in mass with the $f_{0}$-meson.

The $a_{0}$ ($\sigma$) is always a non-strange state and, for $\rho_{sym}=0$, its mass is higher than $\omega_{u}=2M_{u}$. As  the density increases, it 
immediately  becomes a bound state and  degenerates with  its chiral partner $\eta$ ($\pi$) for  densities $\rho_{sym}\simeq 5\rho_0$. 
So, for $\rho_{sym}\gtrsim 5\rho_{0}$ the four bound state mesons ($\pi,\sigma,\,\eta,\,a_{0}$)  become degenerate reflecting the effective restoration of chiral symmetry in the non-strange sector.

Concerning the kaon and its chiral partner $\kappa$ (Fig. \ref{symmass}(h)) they show a clear tendency to get degenerate at high densities, where both mesons are bound states.
We remark that the degeneracy of chiral partners in symmetric quark matter occurs in regions where the mesons are bound states.

Finally, the more significant  difference between  case $g_D=0$ and  the other cases is that the chiral symmetry effective restoration occurs latter, similarly to the situation at non-zero temperature.  

%%%%%%%%%%%%%%%%%%%%%%%%%%%%%%%%%%%%%%%%%%%%%%%%%%%%%%%%%%%%%%%%%%%%%%%%%%%%%%%%%%%%%%%%%%%%%%%%%%%%%%%%%%%%%%%%%%%%%%%%%%%%%%%%%%%%%%%%%%%%%%%%%%%%%%%%%%%%%%%%%%

\section{Results for "neutron" matter in $\beta$ -equilibrium:}

We consider now asymmetric quark matter in weak equilibrium and charge neutrality, supposedly of the same type of the existing in the interior of neutron stars.
To insure this situation, we impose the  following constraints on the chemical potentials and densities of quarks and electrons:
\begin{equation}
\mu_{d}=\mu_{s}=\mu_{u}+\mu_{e} \label{beq}%
\end{equation}
and 
\begin{equation}
\frac{2}{3}\rho_{u}-\frac{1}{3}(\rho_{d}+\rho_{s})-\rho_{e}=0, \label{chneu}%
\end{equation}
with
\begin{equation}
\rho_{i}=\frac{1}{\pi^{2}}(\mu_{i}^{2}-M_{i}^{2})^{3/2}\theta(\mu_{i}%
^{2}-M_{i}^{2})\,\hskip0.2cm \mbox{and}\,\hskip0.2cm \rho_{e}=\mu_{e}^{3}/3\pi^{2}. \label{denes}
\end{equation}

\begin{figure}[t]
\begin{center}
\begin{tabular}
[c]{cc}%
\includegraphics[width=0.53\textwidth]{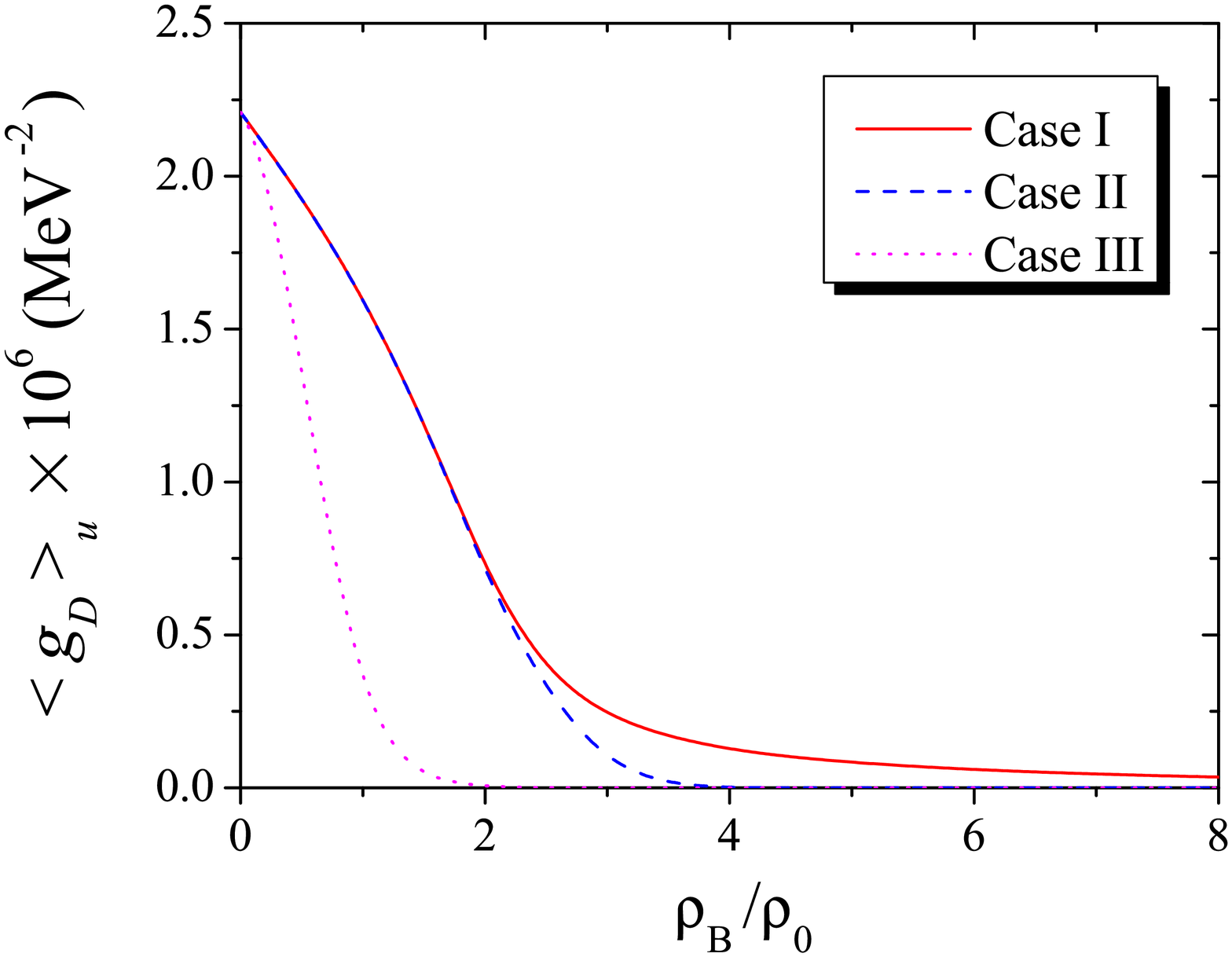} &
\hspace{-1.0cm}\includegraphics[width=0.53\textwidth]{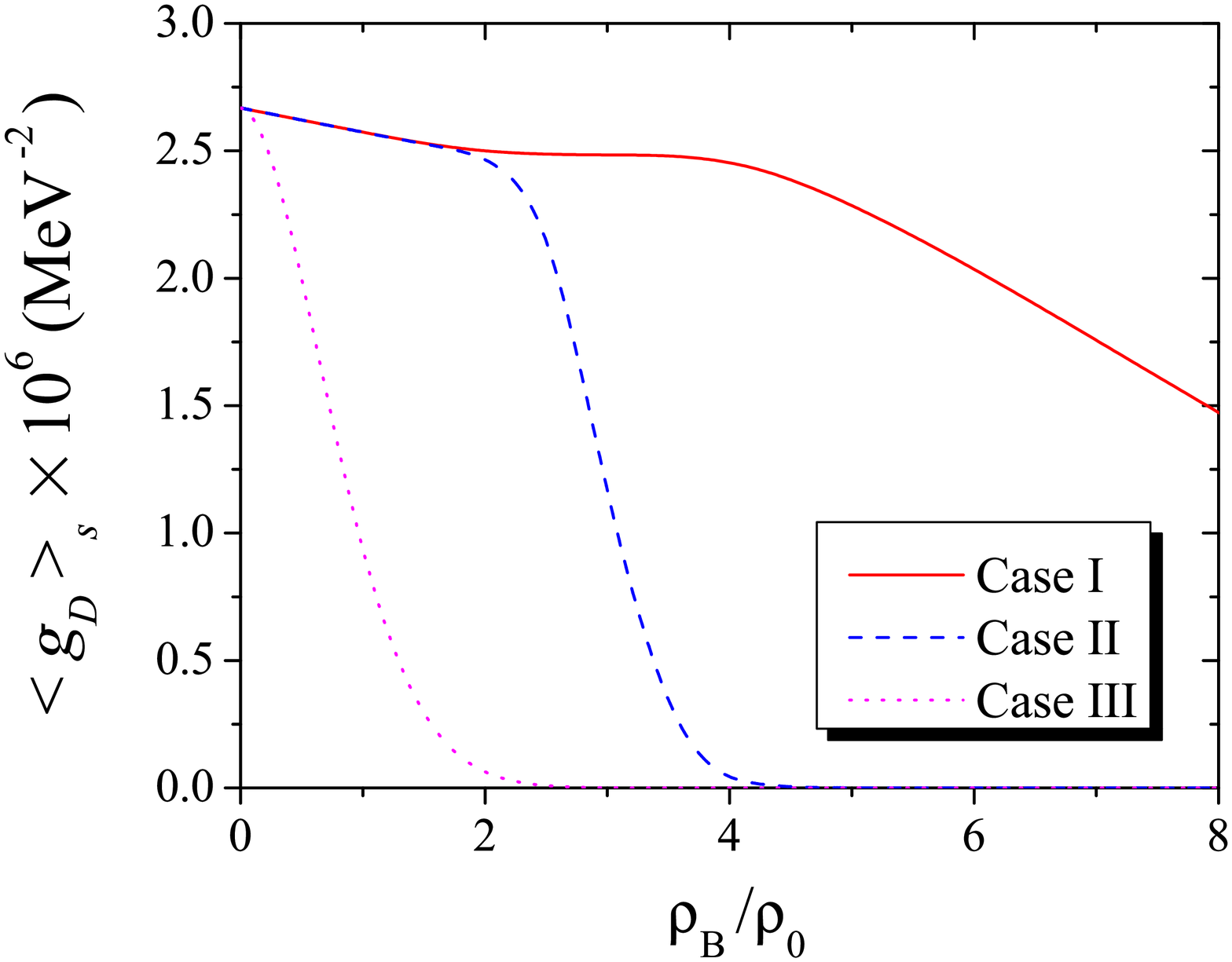}
\end{tabular}
\end{center}
\caption{Density dependence of $\left\langle g_D\right\rangle_u $ (left panel) and $\left\langle g_D\right\rangle_s $ (right panel) for the different cases. }%
\label{gDdens}%
\end{figure}

Similarly to the finite temperature case, and as already explained in the section III,   chiral symmetry is effectively restored  only in the SU(2) sector, in the range of densities  considered, a conclusion  that is independent of the specific form of the dependence on density of the anomaly coefficient, $g_D$.  The effective anomaly coupling,  shown in Fig. \ref{gDdens}, although exhibiting details different from the finite temperature and from the symmetric quark matter cases, are qualitatively similar. 

Let us emphasize some specific aspects on the behavior of the strange quark mass with density. Although in the present case, at low densities, there are no strange quarks in the medium, the mass of the strange quark decreases, although smoothly, due to the  effect of the 't Hooft interaction; eventually it becomes lower than the chemical potential for strange quarks (at $\rho_B \simeq 3.8 \rho_0$). A more pronounced decrease of the strange quark mass is then observed, which is no more due to the anomaly (we can see from Fig. \ref{gDdens} that $\left\langle g_D\right\rangle_{u}$ is already very small) but to the presence of valence strange quarks in the medium (see Eq. (\ref{denes})).

Concerning the meson spectra and the mixing angles, we will show that new aspects also  appear, mainly in the high density region, and will be discussed in the sequel.

%%%%%%%%%%%%%%%%%%%%%%%%%%%%%%%%%%%%
%%%%%%%%%%%%%%%%%%%%%%%%%%%%%%%%%%%%%%%%%%%%%%%%%%%%%%%%%%%%%%%%%%%%%%%%

As it is the only scenario where the flavor symmetry exhibited by the physical  vacuum state  is violated by the weak interaction conditions (\ref{beq}), this implies several consequences:
\begin{itemize}
	\item [(i)] splitting between charge multiplets of pions and kaons;
	\item [(ii)] appearance of low-lying modes above a certain density.
\end{itemize}
  
This leads us to focus on the behavior of all nine pseudoscalar mesons and respective scalar partners, as well as on the chiral partners of the low-lying excitations.
Before we start our discussion we remark the following:

\begin{enumerate}
\item We start by analyzing the chiral asymmetry parameter which is a measure of the violation of the isospin symmetry.

\item We will follow the structure used for the finite temperature and completely symmetric matter which leads to the study  the 
scenarios: Cases I, II and III, with $g_D\neq0$; and the case $g_D=0$.
\end{enumerate}

%%%%%%%%%%%%%%%%%%%%%%%%%%%%%%%%%%%%%%%%%%%%%%%%%%%%%%%%%%%%%%%%%%%%%%%%%%%%%%%%%

\subsection{Chiral asymmetry parameter}
\label{CAP}

\begin{figure}[t]
\begin{center}
\includegraphics[width=0.50\textwidth]{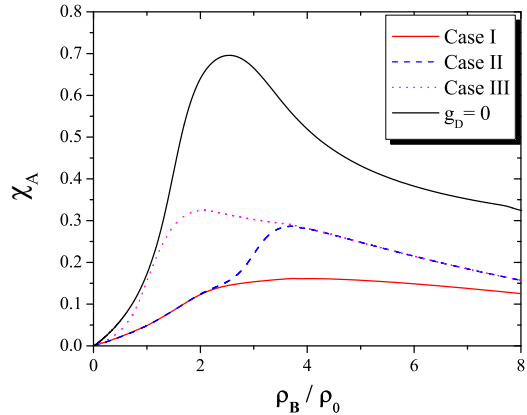}
\end{center}
\caption{Chiral asymmetry parameter as a function of density for the three cases presented in Table II and the case $g_D=0$.}%
\label{chiA}%
\end{figure}

Solving the gap equation (\ref{gap}) one verifies that, in the different cases summarized in Table II, the constituent quark mass $M_d$ decreases slightly more than $M_u$ as the density increases. Bearing in mind a qualitative analysis of the effects of chiral symmetry breaking (restoration) it is useful to plot the isospin asymmetry parameter 
\begin{equation}
\chi_A\,=\,\frac{|M_u-M_d|}{M_u+M_d}, \label{asp}
\end{equation}
as a function of the baryonic density in the several cases under discussion.
As is shown in Fig. \ref{chiA} the chiral asymmetry parameter $\chi_A$ is more significant in the absence of the anomalous coupling constant. 
We remark that the presence of the anomaly in the model has the effect of reducing the isospin asymmetry  in an SU(2) broken system like the neutron matter case. 
The main consequences of this isospin asymmetry of the medium  must be visible  in the behavior of chiral partners.  

%%%%%%%%%%%%%%%%%%%%%%%%%%%%%%%%%%%%
%%%%%%%%%%%%%%%%%%%%%%%%%%%%%%%%%%%%

%%%%%%%%%%%%%%%%%%%%%%%%%%%%%%%%%%%%%%%%%%%%%%%%%%%%%%%%%%%%%%%%%%%%%%%%%%%%%%%%%

\subsection{Explicit chiral symmetry breaking with U$_{A}$(1) anomaly}

\textbf{Case I.} The pseudoscalar sector in neutron matter in $\beta$-equilibrium was extensively studied in \cite{costabig,costaI,costaB}, which corresponds to {Case I} of the present work. Here  we will focus mainly  on the possible degeneracy of chiral partners.

\begin{figure}[t]
\begin{center}
\includegraphics[width=0.75\textwidth]{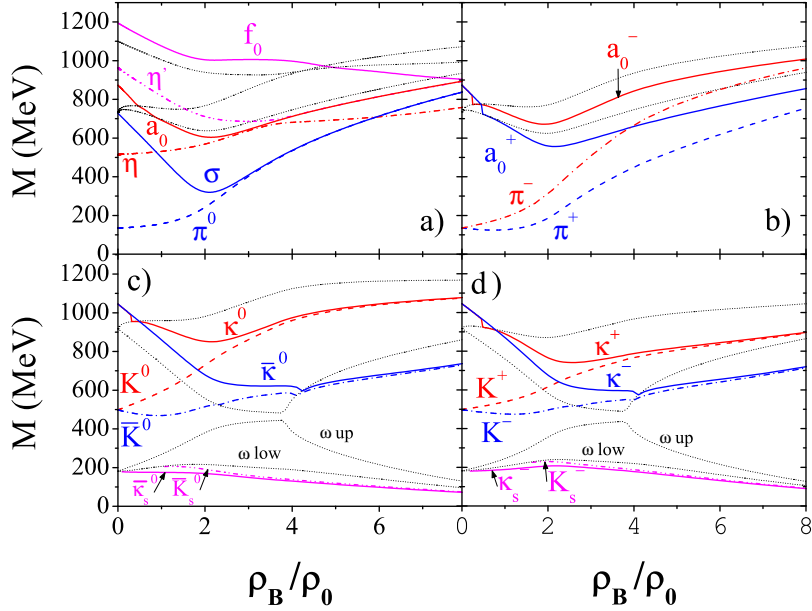}
\end{center}
\caption{{Density dependence of meson masses and of  limits of the Dirac sea continua (dotted lines) defining $q\bar q$ thresholds for the mesons. The low-lying solutions are also included. The anomaly coupling constant is kept constant (Case I).}}%
\label{densnorm}%
\end{figure}

\begin{figure}[t]
\begin{center}
\begin{tabular}
[c]{cc}%
\includegraphics[width=0.53\textwidth]{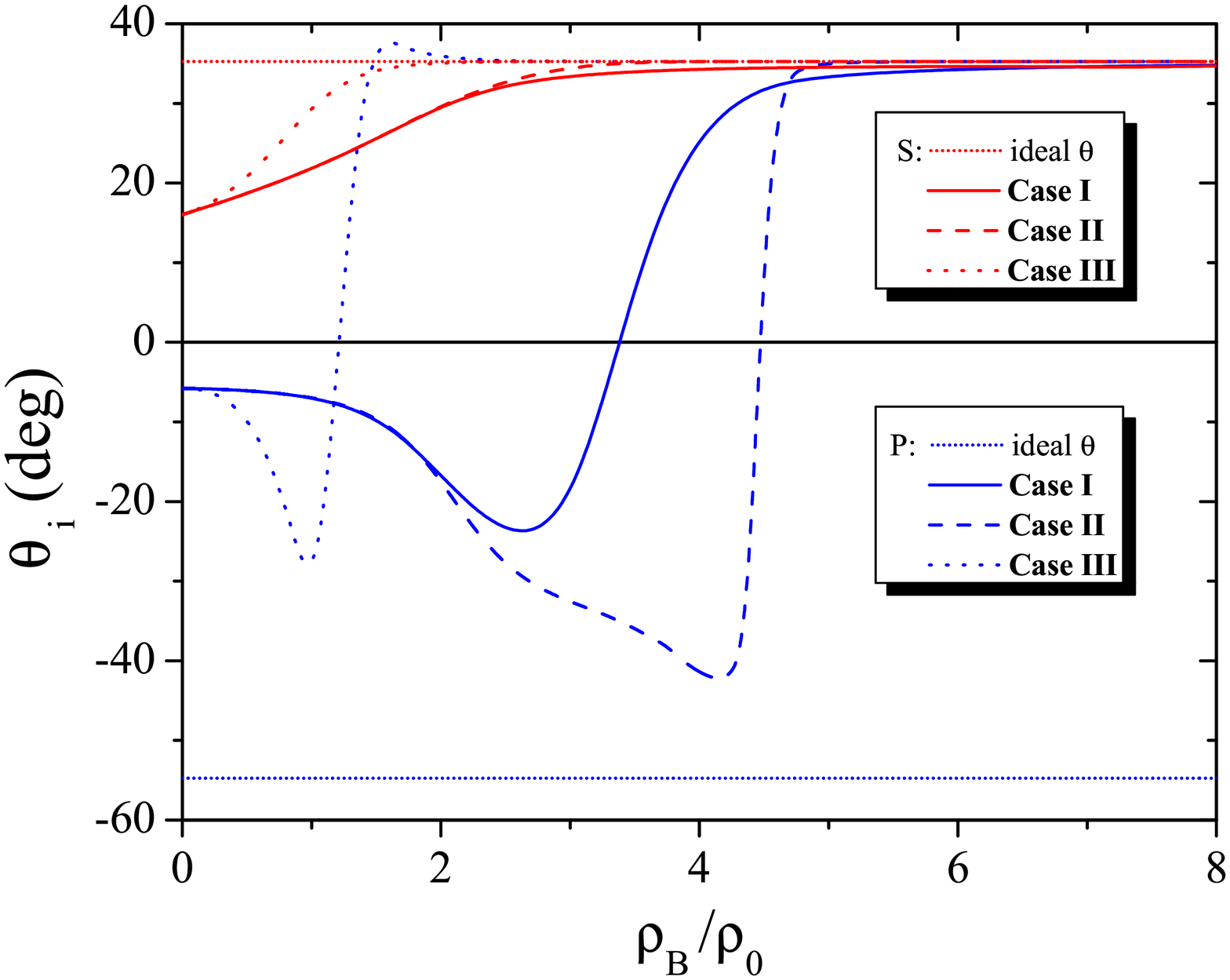} &
\hspace{-1.0cm}\includegraphics[width=0.54\textwidth]{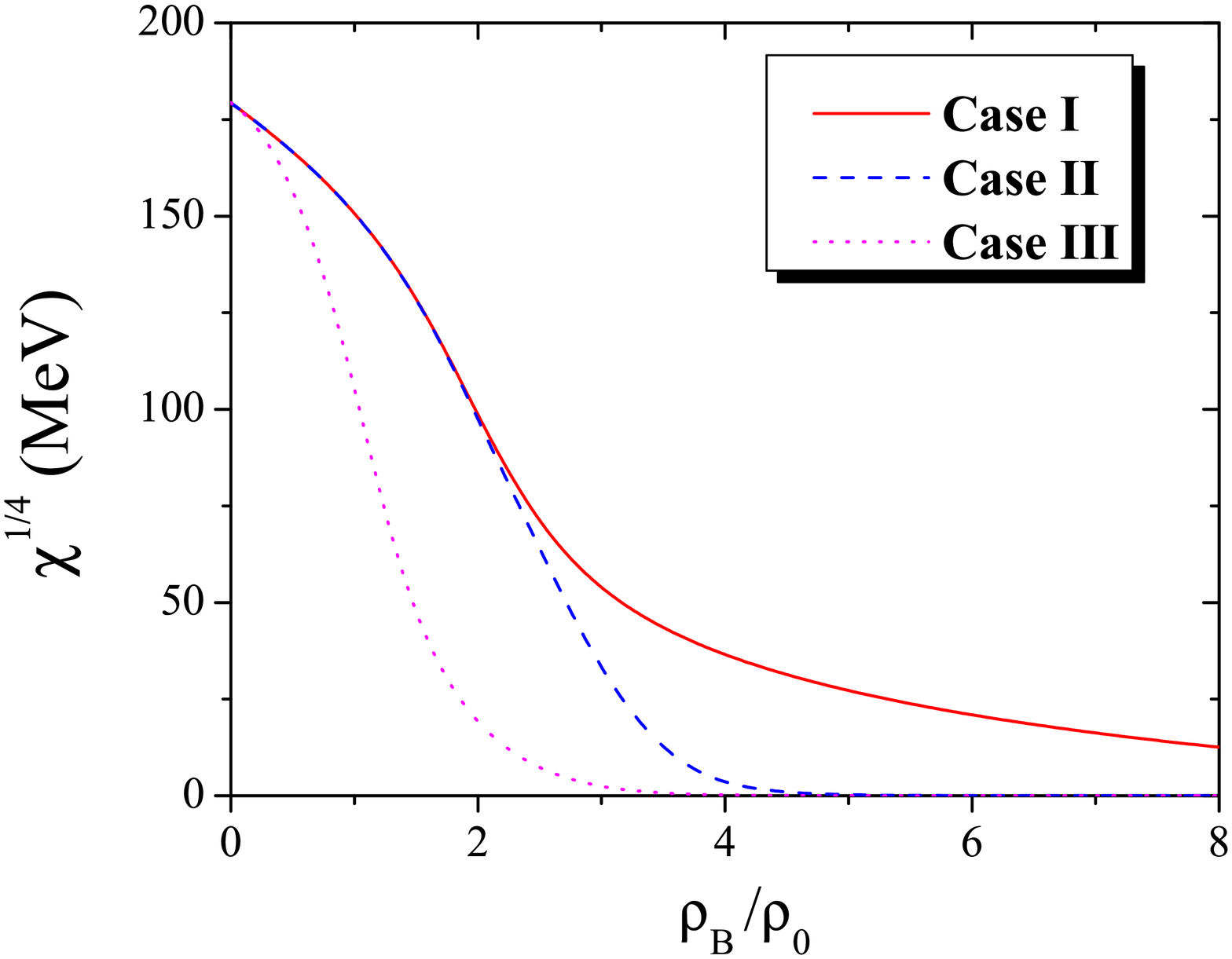}
\end{tabular}
\end{center}
\caption{Left panel: scalar and pseudoscalar mixing angles as a function of density for  the three cases and for the ideal mixing. Right panel: topological susceptibility as a function of density for the three cases.}%
\label{angnorm}%
\end{figure}

In Fig. \ref{densnorm}(a), the meson masses are plotted as functions of the density. The SU(2) chiral partners ($\pi^{0},\sigma$) are always bound states. The pion is a light quark system for all range of densities and the $\sigma$ meson has a strange component at $\rho_{B}=0$ but  never becomes a purely non-strange state because $\theta_{S}$ never reaches $35.264^{\circ}$, the ideal mixing angle (Fig. \ref{angnorm}, {Case I}). As the density increases these mesons become degenerate ($\rho_{B}\gtrsim3\rho_{0}$).
At the same density, the SU(2) chiral partner ($\eta,a_{0}$) is also degenerate. The $\eta$-meson is always a bound state, contrarily to $a_{0}$ that starts as a resonance, once its mass is above the continuum, and becomes a bound state for $\rho_{B}\gtrsim0.5\rho_{0}$. However, the $a_{0}$ mass separates from the $\eta$ mass and goes to degenerate with the $\eta^{\prime}$. To understand this behavior we need to look for the behavior of the mixing angle $\theta_{P}$. From Fig. \ref{angnorm}, {Case I}, we observe that the angle $\theta_{P}$, which starts at $-5.8^{\circ}$, changes sign at $\rho_{B}\simeq 3.5\rho_{0}$ becoming positive and increasing rapidly, which, as will be seen, we interpret as an indication of a  change of identity between $\eta$ and $\eta^\prime$.

%%%%%%%%%%%%%%%%%%%%%%%%%%%%%%%%%%%%
%%%%%%%%%%%%%%%%%%%%%%%%%%%%%%%%%%%%

We remember that up to the density $\rho_B=3.8 \rho_0$ (see end of subsection \ref{CAP}) the $(q\bar q)_s=s\bar s$ content is induced by the mixing effects only. Above this density strange valence  quarks  are present (see Eq. (\ref{denes})) and induce the strange quark mass to decrease faster.  

This behavior  induces changes in the percentage of strange, $(q\bar q)_s=s\bar s$, and non-strange, $(q\bar q)_{\rm ns}=\frac{1}{2}({u}\bar u+{d} \bar d)$, quark content  in $\eta$ and $\eta^\prime$ mesons: at low density, the $\eta^{\prime}$ is more strange than the $\eta$, but the opposite occurs at high density \cite{costaB}.
Then $\eta^{\prime}$ will degenerate in mass with the $a_{0}$-meson that is always a non-strange state. Finally, the $f_{0}$-resonance is always a strange state that shows no tendency to become degenerate with any other meson. 

Now let us comment on the $\pi^{\pm}$ behavior and the respective chiral partners
$a_0^{\,\pm}$ that are plotted in Fig. \ref{densnorm}(b). The $\pi^{\pm}$-mesons are always bound states and their masses increase with density. On the  other side, the $a_0^{\,\pm}$-mesons start as resonances and  become  bound states: the $a_0^{\,-}$ at $\rho_{B}\simeq  0.25\rho_{0}$, and the $a_0^{\,+}$ at $\rho_{B}\simeq 0.5\rho_{0}$. 
However, they never degenerate with the respective pions in the considered range of densities. This is, once again, due to the fact that the chiral symmetry in the strange sector is not  restored, and  the absence of the mechanism of restoration of  the U$_A$(1) symmetry is also relevant in this context.
This will influence the behavior of the $\pi^{\pm}$ and $a_0^{\,\pm}$ mesons through Eqs. (\ref{Ppi}) and (\ref{pci}), respectively, because the quark condensate $\left\langle\bar{s}s\right\rangle$ is still very high (see Fig. \ref {gDdens}  for $ \left\langle g_D\right\rangle_s$).
A different scenario occurs for kaons and their chiral partners: $K^{\pm}$ and $\kappa^{\pm}$ in Fig. \ref{densnorm}(c) and $K^{0}(\bar{K}^{0})$ and $\kappa^{0}(\bar{\kappa}^{0})$ in Fig. \ref{densnorm}(d).
 
As is has already been shown \cite{costabig}, below the lower limit of the Fermi sea continuum of particle-hole excitations, there are low bound states with quantum numbers of $K^{-}\,,\bar{K}^{0}\,$ and $\pi^+$. Here we show that these  low-energy modes, collective particle-hole excitations of the Fermi sea, have corresponding chiral partners. The behavior of the low-energy chiral partners with density is similar to that of the respective high energy modes and does not present meaningful differences in Cases I, II and III. 
This can be seen in Fig. \ref{densnorm}(c) and (d), for ($K^-,\kappa^-$) and ($\bar{K}^0,\bar{\kappa}^{0}$), respectively. 

We also saw in \cite{costabig} that in the present  approach the criterion for the occurrence of kaon condensation is not satisfied since the antikaon masses are always larger than the difference between the chemical potential of strange and non-strange quarks. This conclusion is still valid in the other cases. 

%%%%%%%%%%%%%%%%%%%%%%%%%%%%%%%%%%%%
%%%%%%%%%%%%%%%%%%%%%%%%%%%%%%%%%%%%

\textbf{Case II.} Like in the previous section, we postulate in { Case II} a density dependence of  $\chi$, as a Fermi function, formally similar to the finite temperature case (see Fig. \ref{angnorm}, dashed line in the left panel). Then we can apply this dependence to model the anomalous coupling, allowing the calculation of all observables. 

Analyzing the mixing angles (Fig. \ref{angnorm}, right panel) we observe that the behavior of $\theta_{S}$ is similar to the non-zero temperature and completely symmetric quark matter cases: it starts at $16{{}^\circ}$ and increases up to the ideal mixing angle $35.264{^\circ}$. A different behavior is found for the angle $\theta_{P}$, that changes sign at $\rho_{B}\simeq 4.9\rho_{0}$ ($\simeq 3.50 \rho_0$ in Case I): it starts at $-5.8^{\circ}$ and goes to the ideal mixing angle $35.264{{}^\circ}$, which also leads, by similar reasons as previously, to a change of identity between $\eta$ and $\eta^{\prime}$. 

\begin{figure}[t]
\begin{center}
\includegraphics[width=0.750\textwidth]{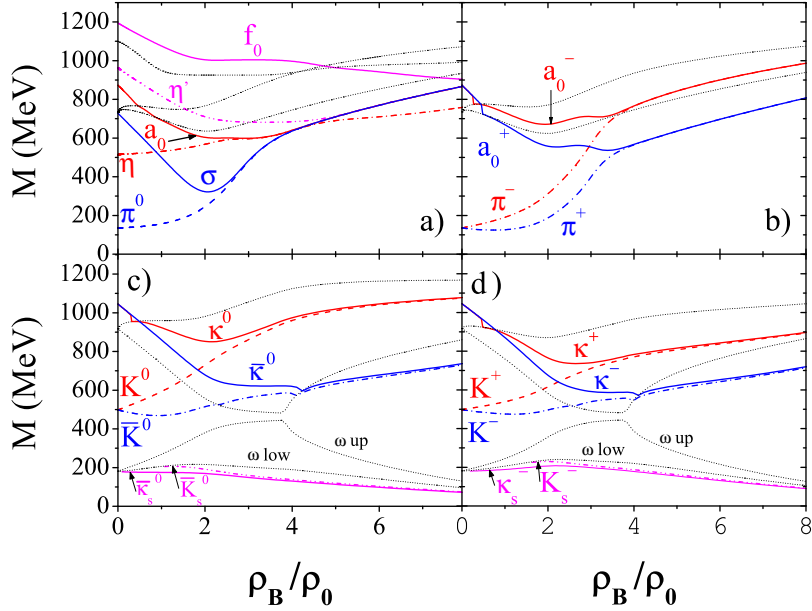}
\end{center}
\caption{Density dependence of meson masses and of  limits of the Dirac sea continua (dotted lines) defining $q\bar q$ thresholds for the mesons. The low-lying solutions are also included. The anomaly coupling  is a Fermi function (Case II).}%
\label{densfermi}%
\end{figure}

The meson masses, as function of the density, are plotted in Fig. \ref{densfermi}(a). The SU(2) chiral partners ($\pi^{0},\sigma$) are now always bound states. The pion is a light quark system for all range of densities and the $\sigma$ meson has a strange component, at $\rho_{B}=0$, but becomes purely non-strange when $\theta_{S}$ goes to $35.264{{}^o}$, at $\rho_{B}\simeq 3\rho_{0}$. At this density the mesons become degenerate.
This behavior is similar to the non-zero temperature case.

The SU(2) chiral partner ($\eta,a_{0}$) becomes degenerate for $4.0\rho_{0}\leq\rho_{B}\leq4.8\rho_{0}$, a region where they are bound states.
In the same range of densities ($\eta,a_{0}$) and ($\pi^{0},\sigma$)
are all degenerate. Suddenly the $\eta$ mass separates from the others
becoming a purely strange state. This is due to the behavior of $\theta_{P}$ that, as already referred,  changes  sign and goes to $35.264^{\circ}$, at $\rho _{B}\simeq 4.9\rho_{0}$. On the other hand, the $\eta^{\prime}$, that starts as an unbound state and becomes bound at $\rho_{B}>3.0\rho_{0}$, turns into a purely light quark system and degenerates with $\pi^{0}$, $\sigma$ and $a_{0}$ mesons. So, the $\eta$ and the $\eta^{\prime}$ also change identities.
Consequently, contrarily to results with temperature, $\pi^{0}$ and $\eta^{\prime}$ are now degenerate. 

Finally we analyze the behavior of charged mesons with density, plotted in Figs. \ref{densfermi} (b) and \ref{densfermi} (d). The figure shows that  the  chiral partners $(\pi^+\,, a_0^{\,+})$ and $(\pi^-\,, a_0^{\,-})$, panel (b), become degenerate for  $\rho_B\simeq 4 \rho_0$; the chiral partners $(K^+\,, \kappa^+)$ and  $(K^-\,, \kappa^-)$, panel (d), and  $(K^0\,, \kappa^0)$ and $(\bar K^0\,, \bar\kappa^0)$, panel (c), do not degenerate in the region of densities considered. 
We notice that,  while the results for $(\pi^\pm\,, a_0^{\,\pm})$ are affected by the dependence of $g_D$ on density, we find no substantial differences for the kaonic mesons, whether  $g_D$ is constant or not. In order to understand this, let us remember that  for the pion and the $a_0$ propagators, the dependence on the anomaly  enters through  the effective coupling $\left\langle g_D\right\rangle_s$ (see Fig. \ref{gDdens}) so, with $g_D$ a decreasing function of the density, this term will affect less and less the meson masses as the density increases. Then, the convergence of the mesons reflects the restoration of  the  U$_A$(1) symmetry.  Since for kaonic mesons the  propagators depend on the anomaly through  the effective  coupling $\left\langle g_D\right\rangle_u$ ($\left\langle g_D\right\rangle_d$), the anomaly has little effect on the kaonic masses, as the density increases, whether $g_D$ is constant or not, due to the strong decrease of the mass of the non-strange quarks. The dominant factor for the calculation of the masses of those mesons is the mass of the strange quark, which, although decreasing, remains always very high.
We can say that the restoration of the axial anomaly does not influence the behavior of kaons and of its chiral partners. 
In addition, we remark that   the chiral asymmetry (\ref{asp}) is always different from zero in neutron matter, even for high densities.    

\begin{figure}[t]
\begin{center}
\includegraphics[width=0.750\textwidth]{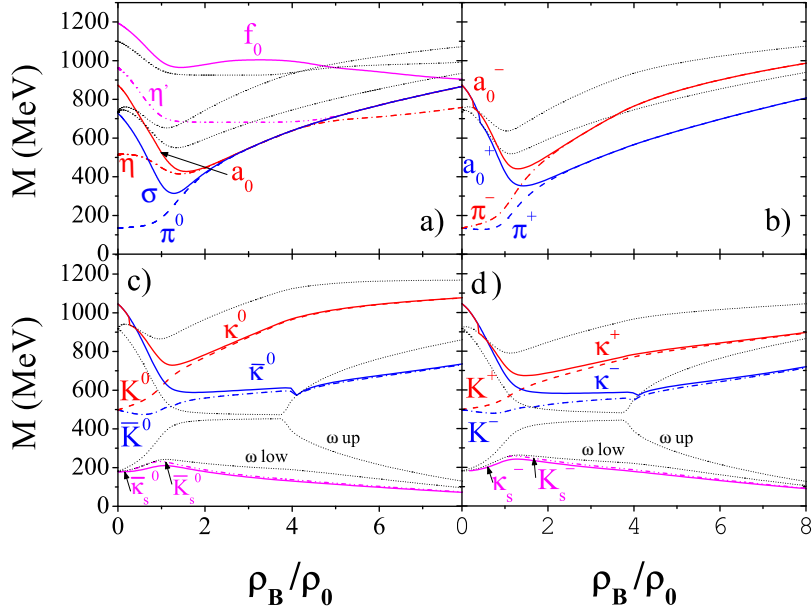}
\end{center}
\caption{Density dependence of meson masses and of  limits of the Dirac sea continua (dotted lines) defining $q\bar q$ thresholds for the mesons. The low-lying solutions are also included. The anomaly coupling  is a decreasing exponential (Case III).}%
\label{densexp}%
\end{figure}

We notice that the convergence between  the different chiral partners always  occurs at densities where the mesons are bound states (see  Figs. \ref{densfermi} and \ref{densexp}), i.e., they are collective excitations defined below the respective $q\bar q$ threshold.

%%%%%%%%%%%%%%%%%%%%%%%%%%%%%%%%%%%%%%%%%%
%%%%%%%%%%%%%%%%%%%%%%%%%%%%%%%%%%%%

\textbf{Case III.} In  this case we postulate the following dependence for $g_{D}:$ $g_{D}(\rho_{B})=g_{D}(0)$exp$[-(\rho_{B}/\rho_{0})^{2}].\ $The topological susceptibility with this dependence is plotted in Fig. \ref{angnorm}, dotted line.
From Fig. \ref{densexp}(a) we see that the density dependence for $g_{D}(\rho_{B})$ used does strengthen the  phase transitions, like in the finite temperature and complete symmetric matter cases. 
The masses of the chiral partners $(\pi^{0},a_{0})$ and $(\eta,\sigma)$ degenerate at very earlier values of the density ($\rho_{B}\simeq 2.5\rho_{0}$), compared with {Case II} (where $\rho_{B}\simeq 4\rho_{0}$). Now the interval where these four mesons are degenerate is bigger: $2.5\rho_{0}\leq\rho_{B}\leq4.8\rho_{0}.$ Then the $\eta$ mass separates from the others becoming a purely strange state and the ($\pi^{0},\,a_{0},\,\eta^{\prime},\,\sigma$) mesons become again degenerate in mass.

In this scenario, $\chi$ (in Fig. \ref{angnorm}, dotted line) goes to zero for
$\rho_{B}\simeq3\rho_{0}$. The behavior of the mixing angles (Fig. \ref{angnorm}, dotted lines) is also qualitatively similar to {Case II}:
$\theta_{S}$ goes to the ideal mixing angles for $\rho_{B}\gtrsim2.5\rho_{0}$
and $\theta_{P}$ also changes sign, however this happens for lower densities,
$\rho_{B}\simeq1.0\rho_{0}$.
In panel (b) of Fig. \ref {densfermi} we verified that the degeneracy of $\pi^{\pm}$ and $a_0^{\,\pm}$ occurs for $\rho_{B}\gtrsim2.5\rho_{0}$ ($\rho_{B}\gtrsim4.0\rho_{0}$ for {Case II}).
In panels (c) and (d) we note a strongest decrease of the $\kappa^{\pm}$ and $\kappa^{0}(\bar{\kappa}^{0})$ masses than in {Case II}. This is the more relevant effect.

%%%%%%%%%%%%%%%%%%%%%%%%%%%%%%%%%%%%
%%%%%%%%%%%%%%%%%%%%%%%%%%%%%%%%%%%%

%%%%%%%%%%%%%%%%%%%%%%%%%%%%%%%%%%%%%%%%%%%%%%%%%%%%%%%%%%%%%%%%%%%%%%%%%%%%%%%%%

\subsection{Explicit chiral symmetry breaking without U$_{A}$(1) anomaly}

\begin{figure}[t]
\begin{center}
\includegraphics[width=0.750\textwidth]{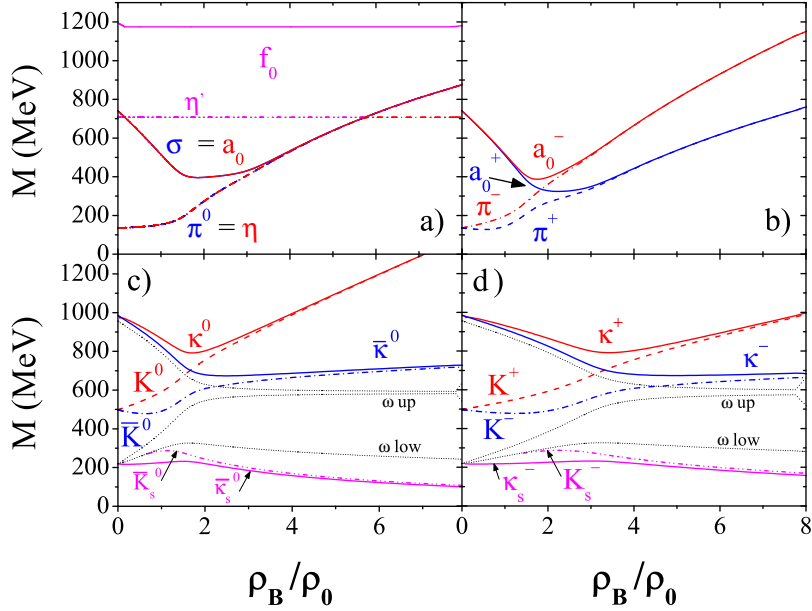}
\end{center}
\caption{Density dependence of meson masses and of  limits of the Dirac sea continua (dotted lines) defining $q\bar q$ thresholds for the mesons. The  low-lying solutions are also included. The axial anomaly is absent  ($g_{D}=0$).}%
\label{densgd0}%
\end{figure}

The absence of mixing effects ($g_D=0$) in the gap equation for the specific environment now considered induces effects that, although in general qualitatively similar to the previous cases studied (finite temperature and symmetric quark matter),  have relevant differences: (i) it is observed a more significant decrease of the constituent quark mass $M_d$ as compared to $M_u$ (the  chiral asymmetric parameter plotted in  Fig. \ref{chiA} reflects this behavior); (ii) the mass of the strange quark remains constant in the range of densities considered, since there are no strange quarks in the medium, due to the fact that $M_s > \mu_s$ (see Eq. \ref{denes}). This two facts will have relevant consequences for the mesonic behavior to be discussed in the sequel. As it can be seen in Fig. \ref{densgd0}(a), and similarly to  the previous situations without anomaly, $\pi^0$ and $\eta$ are degenerate in mass, as well as $a_0$ and $\sigma$ and, as the density increases, the four mesons become degenerate ($\rho_B \simeq 4 \rho_0$). Some meaningfully differences relatives to the other cases with $g_D=0$ appear, however, above $\rho_B \simeq 5.5 \rho_0$. Due to the absence of the anomaly, there are no mixing effects and the mixing angles have, therefore, always ideal values. However, we observe a change of sign of the pseudoscalar angle, $\theta_P$,  at that density ($\theta_P= - 54.736^{\circ}, \mbox{for}\, \rho_B < 5.5 \rho_0\,,\theta_P=35.264^{\circ}, \mbox{for}\, \rho_B > 5.5 \rho_0$) a behavior that seems specific of the type of matter under study. This implies, as usual, that the $\eta$ meson, non strange up to this density, becomes purely strange afterwards, the opposite happening to $\eta'$, that changes the role with $\eta$ from now, being degenerate with $\pi^0\,, a_0\,, \sigma$. A consequence of the strange quark mass remaining constant is that the mesons with only a strangeness content keep their masses constant ($f_0, \eta' (\eta)$).
As in the Cases I, II, and III, the $f_0$-meson shows no tendency to become degenerate with any other meson.

%%%%%%%%%%%%%%%%%%%%%%%%%%%%%%%%%%%%
%%%%%%%%%%%%%%%%%%%%%%%%%%%%%%%%%%%%

In  panel (b) the $a_0^{\,\pm}$ are always bound states and we verify that the degeneracy of ($a_0^{\,-},\pi^-$) and ($a_0^{\,+},\pi^+$) occurs at different baryonic  densities, respectively, $\rho_B\simeq 3.5 \rho_0$ and $\rho_B\simeq 4.1 \rho_0$.
This may indicate the  existence of two separate first-order phase transitions in the non-strange sector, in agreement with the conclusions of \cite{Buballa03}. 

In  panels (c) and (d) we notice a strongest decrease of $\kappa^{0}(\bar{\kappa}^{0})$ masses as compared to those of $\kappa^{\pm}$.
This is due to a more pronounced decrease of $M_d$ with increasing density.
In addition, the splitting between charge multiplets of pions and kaons is always manifest as expected. 

%%%%%%%%%%%%%%%%%%%%%%%%%%%%%%%%%%%%%%%%%%%%%%%%%%%%%%%%%%%%%%%%%%%%%%%%%%%%%%%%%%%%%%%%%%%%%%%%%%%%%%%%%%%%%%%%%%%%%%%%%%%%%%%%%%%%%%%%%%%%%%%%%%%%%%%%%%%%%%%%%%

\section{Summary and conclusions}

In this work we  investigated different patterns of restoration of axial symmetry, in connection with the restoration of chiral symmetry, in a model with explicit breaking of the U$_A$(1) anomaly. The restoration of axial symmetry at non-zero temperature (density)  has been discussed using two different decreasing functions of temperature (density) for the coupling anomaly, $g_D$: one of them is inspired in lattice results (Case II) for the topological susceptibility and the other is a simple exponential function (Case III). This results were compared with the case were  $g_D= {\rm constant} $ for all temperatures and densities.

We verified that in the last  case there is always an amount of U$_A$(1) symmetry breaking in the particle spectrum even when chiral symmetry restoration in the non-strange sector occurs at high temperature (density). 
To complement the information provided  by the effective restoration of axial symmetry, the extreme case $g_D=0$ has also been considered. 
For a more complete understanding of the density effects we  considered two different scenarios  of quark matter: (i) symmetric quark matter; and (ii) neutron matter in $\beta$-equilibrium. 
So, the different patterns of axial symmetry in the vacuum state, with $g_D={\rm constant}$ (Case I), Case II and Case III, and $g_D=0$, have been applied in a hot medium, in symmetric quark matter and in neutron matter.  

Since in all cases  chiral symmetry is explicitly broken by the presence of non-zero current quark mass terms, the chiral symmetry is  realized through parity doubling rather than by massless quarks.
So, the identification of chiral partners and the study of its convergence is the criterion to study the effective restoration of chiral and axial symmetries. An important information is also provided by the mixing angles and we verify that, in the scenario of effective restoration of   axial  symmetry, the mixing angles  converge to the situation of ideal flavor mixing: (i) the $\sigma$ and $\eta$  mesons are pure non-strange $q\bar q$ states, while $f_0$ and $\eta^\prime$ are pure strange $s\bar s$ excitations for  symmetric matter and non-zero temperature cases; (ii) the $\eta$ and $\eta^\prime$ change identities for neutron matter case.

In the conditions of explicit breaking of chiral symmetry (real world) we worked, SU(3) symmetry is not exact and, even in the limiting case $g_D=0$, the strange sector does contribute with significant effects even at high temperature (density) as it is visible in the behavior of $f_0$ and $\eta$ ($\eta^\prime$) mesons.

We can conclude that in Cases II (or III) the U$_A$(1) symmetry is effectively restored above the critical transition temperature of the SU(2) chiral phase transition. But, in the region of temperatures (densities) studied we do not observe signs indicating a full restoration of U(3)$\otimes$U(3)  symmetry as, for instance, the degeneracy of both $a_0$ and $f_0$ mesons with the pion.
In fact, as we work in a real world scenario ($m_u=m_d<<m_s$),  we only observe the return to symmetries of the classical QCD Lagrangian in the non-strange sector.
The dynamics of the system at low temperatures or densities is dominated by quantum effects of both chiral and U$_A$(1) breaking symmetries. This is manifest in the low-lying mesonic spectrum. As the temperature or density increase our model simulates, at least phenomenologically, features of the large hadron mass spectrum. In such systems both chiral and U$_A$(1) symmetries must be restored, which is signaled through a systematical appearance of degenerate chiral and axial partners.

We  started with explicit symmetry breaking in the presence  of the U$_A$(1) anomaly in the vacuum state, with the axial symmetry being effectively restored  by thermal (density) effects. The results are based on a schematic model, however, it includes some of the main ingredients  for a reliable qualitative description of the high temperature or density regime of matter. The anomalous effective interaction vanish under extreme conditions of temperature/density as required by asymptotic freedom of QCD. A more realistic approach, which includes  the enlargement of this behavior to the scalar-pseudoscalar interaction, can be done in the framework of a model with finite range form factors.    

%%%%%%%%%%%%%%%%%%%%%%%%%%%%%%%%%%%%%%%%%%%%%%%%%%%%%%%%%%%%%%%%%%%%%%%%%%%%%%%%%%%%%%%%%%%%%%%%%%%%%%%%%%%%%%%%%%%%%%%%%%%%%%%%%%%%%%%%%%%%%%%%%%%%%%%%%%%%%%%%%%
 
{\large Acknowledgment:}

Work supported by grant SFRH/BD/3296/2000 from FCT (P. Costa), by grant RFBR
03-01-00657, Centro de F\'{\i}sica Te\'orica and GTAE (Yu. Kalinovsky).

%%%%%%%%%%%%%%%%%%%%%%%%%%%%%%%%%%%%%%%%%%%%%%%%%%%%%%%%%%%%%%%%%%%%%%%%%%%%%%%%%%%%%%%%%%%%%%%%%%%%%%%%%%%%%%%%%%%%%%%%%%%%%%%%%%%%%%%%%%%%%%%%%%%%%%%%%%%%%%%%%%

\appendix {}\label{apendice}
\section{}

In this Appendix we present some technical details of the model formalism in the vacuum state and at finite temperature and chemical potential.

%%%%%%%%%%%%%%%%%%%%%%%%%%%%%%%%%%%%%%%%%%%%%%%%%%%%%%%%%%%%%%%%%%%%%%%%%%%%%%%%%

\subsection{Propagators and  polarization operators for pseudoscalar mesons } \label{apend:formalismo}

The effective quark Lagrangian (\ref{lagr_eff}) have been obtained making a contraction of one bilinear $(\bar{q} \lambda^{a} q)$ \cite{RKS,costaI,costaB,costabig,costaD} with  the projectors $S_{ab}\,, P_{ab}$ given by
\begin{eqnarray}
S_{ab} &=& g_S \delta_{ab} + g_D D_{abc}\left\langle\bar{q} \lambda^c q\right\rangle, \label{sab}\\
P_{ab} &=& g_S \delta_{ab} - g_D D_{abc}\left\langle\bar{q} \lambda^c q\right\rangle, \label{pab}
\end{eqnarray}
where $\left\langle\bar{q} \lambda^{c} q\right\rangle$ are vacuum expectation values.
The constants $D_{abc}$ coincide with the SU(3) structure constants $d_{abc}\,\,$ for
$a,b,c =(1,2,\ldots ,8)$ and $D_{0ab}=-\frac{1}{\sqrt{6}}\delta_{ab}$, $D_{000}=\sqrt{\frac{2}{3}}$.

The effective model  Lagrangian (\ref{lagr_eff}) has been  written in a form suitable for the usual bosonization procedure. This can be  done by the integration over the quark fields in the functional integral.
So, the natural degrees of freedom of low-energy QCD in the mesonic sector are achieved. 
It gives the following effective action:  

\begin{align}
W_{eff}[\varphi,\sigma]  &  =-\frac{1}{2}\left(  \sigma^{a}S_{ab}^{-1}%
\sigma^{b}\right)  -\frac{1}{2}\left(  \varphi^{a}P_{ab}^{-1}\varphi
^{b}\right) \nonumber\\
&  -i\mbox{Tr}\,\mbox{ln}\Bigl[i\gamma^{\mu}\partial_{\mu}-\hat{m}%
+\sigma_{a}\lambda^{a}+(i\gamma_{5})(\varphi_{a}\lambda^{a})\Bigr]\,.
\label{action}%
\end{align}

The notation $\mbox{Tr}$ stands for the trace operation over discrete indices
($N_{f}$ and $N_{c}$) and integration over momentum. The fields $\sigma^{a}$
and $\varphi^{a}$ are scalar and pseudoscalar  meson nonets, respectively.

To calculate the meson mass spectrum, we expand the effective action
(\ref{action}) over meson fields. Keeping the pseudoscalar mesons only, we have
the effective meson action
\begin{equation}
W_{eff}^{(2)}[\varphi]=-\frac{1}{2}\varphi^{a}\left[  P_{ab}^{-1}-\Pi_{ab}%
^{P}(P)\right]  \varphi^{b}=-\frac{1}{2}\varphi^{a}(D_{ab}^{P}(P))^{-1}%
\varphi^{b}, \label{act2}%
\end{equation}
with $\Pi_{ab}^{P}(P)$ being the polarization operator, which in the momentum
space has the form
\begin{equation}
\Pi_{ab}^{P}(P)=iN_{c}\int\frac{d^{4}p}{(2\pi)^{4}}\mbox{tr}_{D}\left[
S_{i}(p)(\lambda^{a})_{ij}(i\gamma_{5})S_{j}(p+P)(\lambda^{b})_{ji}%
(i\gamma_{5})\right],  \label{actp}
\end{equation}
where $\mbox{tr}_{D}$ is the trace over Dirac matrices. The expression in
square brackets in (\ref{act2}) is the inverse non-normalized meson
propagator  $(D_{ab}^{P}(P))^{-1}$.

For the non-diagonal mesons $\pi\,,K$, the polarization operator takes the form
\begin{equation}
\Pi_{ij}^{P}(P_{0})=4\left(  (I_{1}^{i}+I_{1}^{j})-[P_{0}^{2}-(M_{i}%
-M_{j})^{2}]\,\,I_{2}^{ij}(P_{0})\right),\label{ppij}
\end{equation}
where  the integrals $I_{1}^{i}$ and $I_{2}^{ij}(P_{0})$ are given by
\begin{equation}
I_{1}^{i}=iN_{c}\int\frac{d^{4}p}{(2\pi)^{4}}\,\frac{1}{p^{2}-M_{i}^{2}}%
=\frac{N_{c}}{4\pi^{2}}\int_{0}^{\Lambda}\frac{\mathtt{p}^{2}d\mathtt{p}%
}{E_{i}}, \label{i1}%
\end{equation}%
\begin{align}
I_{2}^{ij}(P_{0})  &  =iN_{c}\int\frac{d^{4}p}{(2\pi)^{4}}\,\frac{1}%
{(p^{2}-M_{i}^{2})((p+P_{0})^{2}-M_{j}^{2})}\nonumber\label{i2}\\
&  =\frac{N_{c}}{4\pi^{2}}\int_{0}^{\Lambda}\frac{\mathtt{p}^{2}d\mathtt{p}%
}{E_{i}E_{j}}\,\,\,\frac{E_{i}+E_{j}}{P_{0}^{2}-(E_{i}+E_{j})^{2}}\,,
\end{align}
where $E_{i,j}=\sqrt{\mathtt{p}^{2}+M_{i,j}^{2}}$ is the quark energy. To
regularize the integrals we introduce the 3-dimensional cut-off parameter
$\Lambda$. When $P_{0}>M_{i}+M_{j}$ it is necessary to take into account the
imaginary part of the second integral. It may be found, with help of the
$i\epsilon$ -prescription $P_{0}^{2}\rightarrow P_{0}^{2}-i\epsilon$. Using 
\begin{equation}
\lim_{\epsilon\rightarrow0^{+}}\frac{1}{y-i\epsilon}={\mathcal{P}}\frac{1}%
{y}+i\pi\delta(y)
\end{equation}
we obtain the integral
\begin{equation}
I_{2}^{ij}(P_{0})=\frac{N_{c}}{4\pi^{2}}{\mathcal{P}}\int_{0}^{\Lambda}%
\frac{\mathtt{p}^{2}d\mathtt{p}}{E_{i}E_{j}}\,\,\frac{E_{i}+E_{j}}{P_{0}%
^{2}-(E_{i}+E_{j})^{2}}+i\frac{N_c}{16\pi}\,\frac{p^{\ast}}{(E_{i}^{\ast}%
+E_{j}^{\ast})}, \label{ima}%
\end{equation}
with the momentum: $p^{\ast}=\sqrt{(P_{0}^{2}-(M_{i}-M_{j})^{2})(P_{0}%
^{2}-(M_{i}+M_{j})^{2})}/2P_{0}$ and the energy: $E_{i,j}^{\ast}%
=\sqrt{(p^{\ast})^{2}+M_{i,j}^{2}}$.

To consider the diagonal mesons $\pi^{0}$, $\eta$ and $\eta^{\prime}$ we take
into account the matrix structure of the propagator in (\ref{act2}). In the
basis of $\pi^{0}-\eta-\eta^{\prime}$ system we write the projector $P_{ab}$
and the polarization operator $\Pi_{ab}^{P}$ as matrices:
\begin{equation}
{P}_{ab}=\left(
\begin{array}
[c]{ccc}%
P_{33} & P_{30} & P_{38}\\
P_{03} & P_{00} & P_{08}\\
P_{83} & P_{80} & P_{88}%
\end{array}
\right)  \,\,\,\,\,\,\mbox{and}\,\,\,\,\,\,{\Pi}_{ab}^{P}=\left(
\begin{array}
[c]{ccc}%
\Pi_{33}^{P} & \Pi_{30}^{P} & \Pi_{38}^{P}\\
\Pi_{03}^{P} & \Pi_{00}^{P} & \Pi_{08}^{P}\\
\Pi_{83}^{P} & \Pi_{80}^{P} & \Pi_{88}^{P}%
\end{array}
\right). \label{Pab}%
\end{equation}

The non-diagonal matrix elements $P_{30}=\frac{1}{\sqrt{6}}g_{D}(\left\langle\bar{q}%
_{u}\,q_{u}\right\rangle-\left\langle\bar{q}_{d}\,q_{d}\right\rangle)$, $P_{38}=-\frac{1}{\sqrt{3}}g_{D}(\left\langle\bar
{q}_{u}\,q_{u}\right\rangle-\left\langle\bar{q}_{d}\,q_{d}\right\rangle)$, $\Pi_{30}=\sqrt{2/3}[\Pi_{uu}%
^{P}(P_{0})-\Pi_{dd}^{P}(P_{0})]$ and $\Pi_{38}=1/\sqrt{3}[\Pi_{uu}^{P}%
(P_{0})-\Pi_{dd}^{P}(P_{0})]$ correspond to $\pi^{0}-\eta$ and $\pi^{0}%
-\eta^{\prime}$ mixing. In the case $\left\langle\bar{q}_{u}\,q_{u}\right\rangle=\left\langle\bar{q}_{d}%
\,q_{d}\right\rangle$, the $\pi^{0}$ is decoupled from the $\eta-\eta^{\prime}$ system and
the preceding matrices have the non-vanishing elements:%

\begin{align}
P_{33}  &  =g_{S}+g_{D}\left\langle\bar{q}_{s}\,q_{s}\right\rangle,\\
P_{00}  &  =g_{S}-\frac{2}{3}g_{D}\left(  \left\langle\bar{q}_{u}\,q_{u}\right\rangle+\left\langle\bar{q}%
_{d}\,q_{d}\right\rangle+\left\langle\bar{q}_{s}\,q_{s}\right\rangle\right)  ,\\
P_{88}  &  =g_{S}+\frac{1}{3}g_{D}\left(  2\left\langle\bar{q}_{u}\,q_{u}\right\rangle+2\left\langle\bar{q}%
_{d}\,q_{d}\right\rangle-\left\langle\bar{q}_{s}\,q_{s}\right\rangle\right)  ,\\
P_{08}  &  =P_{80}=\frac{1}{3\sqrt{2}}g_{D}\left(  \left\langle\bar{q}_{u}\,q_{u}%
\right\rangle+\left\langle\bar{q}_{d}\,q_{d}\right\rangle-2\left\langle\bar{q}_{s}\,q_{s}\right\rangle\right)  ,
\end{align}
and
\begin{align}
\Pi_{00}^{P}(P_{0})  &  =\frac{2}{3}\left[  \Pi_{uu}^{P}(P_{0})+\Pi_{dd}%
^{P}(P_{0})+\Pi_{ss}^{P}(P_{0})\right]  ,\\
\Pi_{88}^{P}(P_{0})  &  =\frac{1}{3}\left[  \Pi_{uu}^{P}(P_{0})+\Pi_{dd}%
^{P}(P_{0})+4\Pi_{ss}^{P}(P_{0})\right]  ,\\
\Pi_{08}^{P}(P_{0})  &  =\Pi_{80}^{P}(P_{0})=\frac{\sqrt{2}}{3}\left[
\Pi_{uu}^{P}(P_{0})+\Pi_{dd}^{P}(P_{0})-2\Pi_{ss}^{P}(P_{0})\right]  ,
\end{align}
where
\begin{equation}
\Pi_{ii}^{P}(P_{0})=4(2I_{1}^{i}-P_{0}^{2}I_{2}^{ii}(P_{0})).
\end{equation}

The procedure to describe scalar mesons is analogous.
We present below the most relevant steps.

To calculate the meson mass spectrum, we expand the effective action
(\ref{action}) over meson fields. Keeping  now the scalar mesons only, we have
the effective meson action
\begin{equation}
W_{eff}^{(2)}[\sigma]=-\frac{1}{2}\sigma^{a}\left[  S_{ab}^{-1}-\Pi_{ab}%
^{S}(P)\right]  \sigma^{b}=-\frac{1}{2}\sigma^{a}({D}_{ab}^{S}(P))^{-1}%
\sigma^{b}, \label{accao2}%
\end{equation}
with $\Pi_{ab}^{S}(P)$ being the polarization operator, which in the momentum
space has the form of (\ref {actp}) with ($i\gamma_5$) substituted by the identity matrix.

The polarization operator associated with the non-diagonal mesons ($a_0,\,\sigma,\,f_0$) has the form
\begin{equation}
\Pi_{ij}^{S}(P_{0})=4\left(  (I_{1}^{i}+I_{1}^{j})+[P_{0}^{2}-(M_{i}{}%
^{2}+M_{j}^{2})]\,\,I_{2}^{ij}(P_{0})\right)  \,.\label{ssij}
\end{equation}

To consider the diagonal mesons $a_0^{\,0}$, $\sigma$ and
$f_0$ we take into account the matrix structure of the
propagator in (\ref{accao2}). In the basis of $a_0^{\,0}-\sigma-f_0$ system we write the projector $S_{ab}$ and the
polarization operator $\Pi_{ab}^{S}$ as matrices:
\begin{equation}
{S}_{ab}=\left(
\begin{array}
[c]{ccc}%
S_{33} & S_{30} & S_{38}\\
S_{03} & S_{00} & S_{08}\\
S_{83} & S_{80} & S_{88}%
\end{array}
\right)  \,\,\,\,\,\,\mbox{and}\,\,\,\,\,\,{\Pi}_{ab}^{S}=\left(
\begin{array}
[c]{ccc}%
\Pi_{33}^{S} & \Pi_{30}^{S} & \Pi_{38}^{S}\\
\Pi_{03}^{S} & \Pi_{00}^{S} & \Pi_{08}^{S}\\
\Pi_{83}^{S} & \Pi_{80}^{S} & \Pi_{88}^{S}%
\end{array}
\right)  .
\end{equation}
In the case $\left\langle\bar{q}_{u}\,q_{u}\right\rangle=\left\langle\bar{q}_{d}\,q_{d}\right\rangle$ the preceding form of
the matrices is reduced to
\begin{equation}
{S}_{ab}\rightarrow\left(
\begin{array}
[c]{cc}%
S_{33} & 0\\
0 & \bar{S}_{ab}%
\end{array}
\right)  \,\,\,\,\,\,\mbox{and}\,\,\,\,\,\,{\Pi}_{ab}^{P}\rightarrow\left(
\begin{array}
[c]{cc}%
\Pi_{33}^{S} & 0\\
0 & \bar{\Pi}_{ab}^{S}%
\end{array}
\right)  ,
\end{equation}
with
\begin{align}
S_{33}  &  =g_{S}-g_{D}\left\langle\bar{q}_{s}\,q_{s}\right\rangle,\\
S_{00}  &  =g_{S}+\frac{2}{3}g_{D}\left(  \left\langle\bar{q}_{u}\,q_{u}\right\rangle+\left\langle\bar{q}%
_{d}\,q_{d}\right\rangle+\left\langle\bar{q}_{s}\,q_{s}\right\rangle\right)  ,\\
S_{88}  &  =g_{S}-\frac{1}{3}g_{D}\left(  2\left\langle\bar{q}_{u}\,q_{u}\right\rangle+2\left\langle\bar{q}%
_{d}\,q_{d}\right\rangle-\left\langle\bar{q}_{s}\,q_{s}\right\rangle\right)  ,\\
S_{08}  &  =S_{80}=-\frac{1}{3\sqrt{2}}g_{D}\left(  \left\langle\bar{q}_{u}\,q_{u}%
\right\rangle+\left\langle\bar{q}_{d}\,q_{d}\right\rangle-2\left\langle\bar{q}_{s}\,q_{s}\right\rangle\right)  .
\end{align}
Analogously, we get
\begin{align}
\Pi_{00}^{S}(P_{0})  &  =\frac{2}{3}\left[  \Pi_{uu}^{S}(P_{0})+\Pi_{dd}%
^{S}(P_{0})+\Pi_{ss}^{S}(P_{0})\right]  ,\\
\Pi_{88}^{S}(P_{0})  &  =\frac{1}{3}\left[  \Pi_{uu}^{S}(P_{0})+\Pi_{dd}%
^{S}(P_{0})+4\Pi_{ss}^{S}(P_{0})\right]  ,\\
\Pi_{08}^{S}(P_{0})  &  =\Pi_{80}^{S}(P_{0})=\frac{\sqrt{2}}{3}\left[
\Pi_{uu}^{S}(P_{0})+\Pi_{dd}^{S}(P_{0})-2\Pi_{ss}^{S}(P_{0})\right]  ,
\end{align}
where
\begin{equation}
\Pi_{ii}^{S}(P_{0})=4(2I_{1}^{i}+[P_{0}^{2}-4M_{i}^{2}]I_{2}^{ii}(P_{0})).
\end{equation}

We also obtain
\begin{equation}
D_{\sigma}^{-1}=\left(  \mathcal{A}+\mathcal{C}\right)  -\sqrt
{(\mathcal{C}-\mathcal{A})^{2}+4\mathcal{B}^{2}}%
\end{equation}
and
\begin{equation}
D_{f_0}^{-1}=\left(  \mathcal{A}+\mathcal{C}\right)
+\sqrt{(\mathcal{C}-\mathcal{A})^{2}+4\mathcal{B}^{2}}\,,
\end{equation}
where the expressions for $ \mathcal{A}$, $ \mathcal{B}$ and $ \mathcal{C}$ are formally analogous to those for pseudoscalars.

The masses of the $\sigma$ and $f_0$ meson can now be determined by the conditions $D_{\sigma}^{-1}(M_{\sigma},\mathbf{0})=0$ and $D_{f_0}^{-1}(M_{f_0},\mathbf{0})=0\,$.

%%%%%%%%%%%%%%%%%%%%%%%%%%%%%%%%%%%%%%%%%%%%%%%%%%%%%%%%%%%%%%%%%%%%%%%%%%%%%%%%%

\subsection{Model formalism at finite temperature and chemical potential} \label{apend:Matsubara}

The NJL model can be generalized to the finite temperature and
chemical potential case.
It can be done by the substitution \cite{kapusta}
\begin{align}
\label{subs}\int\frac{d^{4} p}{(2\pi)^{4}} \longrightarrow\frac{1}{-i \beta}
\int\frac{d^{3} {p}}{(2\pi)^{3}} \sum_{n} \, ,
\end{align}
where $\beta= 1/T$, $T$ is the temperature and the sum is done over Matsubara
frequencies $\omega_{n}=(2n+1)\pi T$, $n=0,\pm1, \pm2, \ldots$, so that $p_{0}
\longrightarrow i\omega_{n} + \mu$ with a chemical potential $\mu$. 
Instead of integration over $p_{0}$ we have now the sum over Matsubara frequencies which can be evaluated

\begin{align}
- \frac{1}{\beta} \sum_{n} h(\omega_{n})  &  = \sum_{\mbox{Re} z_{m} \neq0}
\biggl[\left(  1-f(z_{m}) \right)  \mbox{Res} [ h(\omega_{n}),z_{m}
]\nonumber\\
&  + \bar{f}(z_{m}) \mbox{Res} [ \bar{h}(\omega_{n}),z_{m} ] \biggr] \, ,
\end{align}
where $f(z)$ and $\bar{f}(z)$ are the Fermi distribution functions for quarks
and antiquarks:
\begin{align}
f(z) = \frac{1}{1+ e^{\beta(z-u)}},\,\,\,\bar{f}(z) = \frac{1}{1+
e^{\beta(z+u)}}.
\end{align}
As $1-\bar{f}(z)=f(-z)$, we introduce, for convenience, the Fermi distribution
functions for the positive (negative) energy state of the $i$th quark:
\begin{align}
n_{i}^{\pm}= f_{i}(\pm E_{i}) = \frac{1}{1+ e^{\pm\beta(E_{i}\mp\mu_{i})}}.
\end{align}

At finite temperature the integral $I_{1}^{i}$ (\ref{i1}) takes the form 

\begin{align}
\label{firstt}I_{1}^{i} (T\,,\mu_{i} )= - \frac{N_{c}}{4\pi^{2}} \int
\frac{\mathtt{p}^{2} d\mathtt{p}}{E_{i}} \left(  n^{+}_{i} - n^{-}_{i}
\right)  .
\end{align}

The integral $I_{2}^{ij}(P)$ depends now on the temperature $T$ and two chemical potentials $\mu_{i}, \mu_{j}$ which are appropriated to quark flavors
\begin{align}
I_{2}^{ij} (P_{0},T,\mu_{i},\mu_{j})  &  = - N_{c} \int\frac{d^{3}\mathbf{p}%
}{(2\pi)^{3}} \Biggl[ \frac{1}{2E_{i}} \frac{1}{(E_{i}+P_{0}- (\mu_{i}-\mu
_{j}))^{2}-E_{j}^{2}} \,\, n^{+}_{i}\nonumber\\
&  \hspace*{1.3cm} - \frac{1}{2E_{i}} \frac{1}{(E_{i}-P_{0}+ (\mu_{i}-\mu
_{j}))^{2}-E_{j}^{2}} \,\, n^{-}_{i}\nonumber\\
&  \hspace*{1.3cm}+ \frac{1}{2E_{j}}\frac{1}{(E_{j}-P_{0}+ (\mu_{i}-\mu
_{j}))^{2}-E_{i}^{2}} \,\, n^{+}_{j}\nonumber\\
&  \hspace*{1.3cm}- \frac{1}{2E_{j}}\frac{1}{(E_{j}+P_{0}- (\mu_{i}-\mu
_{j}))^{2}-E_{i}^{2}} \,\, n^{-}_{j} \Biggr] \, .
\end{align}
For the case $i=j$, with imaginary part, we have the expression
\begin{align}
\label{sint}I_{2}^{ii}(P_{0}, T, \mu_{i}) =  &  - \frac{N_{c}}{2\pi^{2}}
{\mathcal{P}} \int\frac{\mathtt{p}^{2} d \mathtt{p}}{E_{i}} \,\, \frac
{1}{P_{0}^{2}-4 E_{i}^{2}} \left(  n^{+}_{i} - n^{-}_{i}\right) \nonumber\\
&  - i \frac{N_{c}}{4\pi} \sqrt{ 1- \frac{4 M_{i}^{2}}{P_{0}^{2}} } \left(
n^{+}_{i}(\frac{P_{0}}{2}) - n^{-}_{i} (\frac{P_{0}}{2})\right)  \,.
\end{align}

Having these integrals as functions of the temperature and chemical potentials, we can investigate the meson properties in hot/dense matter.

%%%%%%%%%%%%%%%%%%%%%%%%%%%%%%%%%%%%%%%%%%%%%%%%%%%%%%%%%%%%%%%%%%%%%%%%%%%%%%%%%

\subsection{Topological susceptibility} \label{apend:TS}

The topological susceptibility is given by
\begin{equation}
 \chi(k^2)=\int{\rm d}^4x\;{\rm e}^{-ikx}\langle0|
 TQ(x)Q(0)|0\rangle_{\rm connected},
\end{equation}
where $Q(x)$ is the topological charge density. 
The general expression for $\chi$ in NJL model has been obtained in \cite{Ohta}

\begin{eqnarray}
\chi &=& 4g_D^2\Bigg[
 2\Pi^{P}_{uu}(0)\left\langle\bar q_u\,q_u\right\rangle^2\left\langle\bar q_s\, q_s\right\rangle^2+
 \Pi^{P}_{ss}(0)\left\langle\bar q_u\,q_u\right\rangle^4			 									\nonumber\\
&+&\Bigg\{\frac{1}{\sqrt{3}}\left\langle\bar q_u\,q_u\right\rangle
 \left(\left\langle\bar q_s\,q_s\right\rangle-\left\langle\bar q_u\,q_u\right\rangle\right)
 {\Pi^{P}_{88}\choose \Pi^{P}_{80}}^{\text{t}} 							\nonumber\\
&+&\frac{1}{\sqrt{6}}\left\langle\bar q_u\,q_u\right\rangle
 \left(2\left\langle\bar q_s\,q_s\right\rangle+\left\langle\bar q_u\,q_u\right\rangle\right)
 {\Pi^{P}_{08}\choose \Pi^{P}_{00}}^{\text{t}}\Bigg\} \nonumber\\
&\times& 2\hat{K}\left(1-2\hat{\Pi} \hat{K}\right)^{-1} \nonumber\\
&\times& \Bigg\{\frac{1}{\sqrt{3}}\left\langle\bar q_u\,q_u\right\rangle
 \left(\left\langle\bar q_s\,q_s\right\rangle-\left\langle\bar q_u\,q_u\right\rangle\right)
 {\Pi^{P}_{88}\choose\Pi^{P}_{08}}													\nonumber\\
&+&\frac{1}{\sqrt{6}}\left\langle\bar q_u\,q_u\right\rangle
 \left(2\left\langle\bar q_s\,q_s\right\rangle+\left\langle\bar q_u\,q_u\right\rangle\right)
 {\Pi^{P}_{80}\choose\Pi^{P}_{00}}\Bigg\}\Bigg].\label{susc}
\end{eqnarray}

%%%%%%%%%%%%%%%%%%%%%%%%%%%%%%%%%%%%%%%%%

\end{document}